\documentclass[preprint]{aastex}
\shorttitle{21cm Reionization Sensitivity}
\shortauthors{Parsons, et al.}

\usepackage{amsmath}
\usepackage{graphicx}
\usepackage{natbib}
\begin{document}
\title{A Sensitivity and Array-Configuration Study for Measuring the Power Spectrum of $21$cm Emission from Reionization}

\author{Aaron Parsons\altaffilmark{1}, Jonathan Pober\altaffilmark{1}, Matthew McQuinn\altaffilmark{1}, Daniel Jacobs\altaffilmark{2}, and James Aguirre\altaffilmark{2}}

\altaffiltext{1}{University of California, Berkeley}
\altaffiltext{2}{University of Pennsylvania}


\begin{abstract}
Telescopes aiming to measure 21cm emission from the Epoch of Reionization must
toe a careful line, balancing the need for raw sensitivity against the
stringent calibration requirements for removing bright foregrounds.  It is
unclear what the optimal design is for achieving both of these goals.  
Via a 
pedagogical derivation of an interferometer's response to
the power spectrum of 21cm reionization fluctuations, we show that even
under optimistic scenarios, first-generation arrays will yield 
low-SNR detections, and that different compact array configurations can substantially alter sensitivity.  We explore the sensitivity gains 
of array
configurations that yield high redundancy in the $uv$-plane -- configurations
that have been largely ignored since the advent of self-calibration for
high-dynamic-range imaging.  We first introduce a mathematical
framework to generate optimal minimum-redundancy configurations for imaging.  We contrast
the sensitivity of such configurations with high-redundancy configurations,
finding that high-redundancy configurations can improve power-spectrum sensitivity
by more than an order of magnitude.  We
explore how high-redundancy array configurations can be tuned to various
angular scales, enabling array sensitivity to be directed away from regions of
the $uv$-plane (such as the origin) where foregrounds are brighter
and where instrumental systematics are more problematic.   We
demonstrate that a 132-antenna deployment of the Precision Array for Probing
the Epoch of Reionization (PAPER) observing for 120 days in a high-redundancy
configuration will, under ideal conditions, have the requisite sensitivity to detect the power spectrum
of the 21cm signal from reionization at a $3\sigma$ level at $k<0.25h\ {\rm
Mpc}^{-1}$ in a bin of $\Delta{\rm ln}~k=1$.  We discuss the tradeoffs of
low- versus high-redundancy configurations.
\end{abstract}


\section{Introduction}


The Epoch of Reionization (EoR) --- the rapid ionization of the majority of the
hydrogen in the universe by light from the first stars and black
holes --- is the most recent phase transition in the state of the baryons in our universe, 
and yet it still remains
largely unexplored.  Observations of redshifted emission from the 21cm
hyperfine transition of neutral hydrogen have the potential to provide unrivaled
detail about this epoch \citep{field1958,madau_et_al1997,furlanetto_et_al2006}.
Variations in this signal versus redshift and direction allow the
reconstruction of a three-dimensional map of the evolution of the ionization state of the hydrogen.  However, reaching the sensitivity to image the structures during reionization
will require an instrument with roughly a square kilometer of collecting area
\citep{mcquinn_et_al2006}.  As a result, first-generation radio telescopes targeting
reionization either aim to measure the global temperature change of 21cm emission during EoR (a task which is not sensitivity-limited), as with
the Compact Reionization Experiment (CoRE) and
the Experiment to Detect the Global EOR Signature (EDGES; \citealt{bowman_rogers2010}),
or instead aim for a statistical detection of the 21cm fluctuations generated by reionization, as with
the Giant Metre-wave Radio Telescope (GMRT;
\citealt{pen_et_al2009})\footnote{\url{http://gmrt.ncra.tifr.res.in/}},
the LOw Frequency ARray (LOFAR; \citealt{rottgering_et_al2006})\footnote{\url{http://www.lofar.org/}}, 
the Murchison Widefield Array (MWA; \citealt{lonsdale_et_al2009})\footnote{\url{http://www.mwatelescope.org/}}, and the Donald C. Backer Precision
Array for Probing the Epoch of Reionization (PAPER; \citealt{parsons_et_al2010})\footnote{\url{http://eor.berkeley.edu/}}.
A detection of the EoR by the first generation of instruments would
establish low-frequency radio astronomy as a powerful probe of
reionization and of the high-redshift universe.

Removing foregrounds that are 
orders of magnitude brighter than the signal and obtaining the
requisite sensitivity are the primary
concerns that influence the design of all 21cm instruments \citep{parsons_et_al2010,bowman_et_al2009,paciga_et_al2010}.
Astrophysical foregrounds that interfere
with the direct detection of a 21cm EoR signature arise primarily from
 galactic and extragalactic synchrotron and free-free emission.  With the
exception of leakage terms from Faraday rotated polarized galactic synchrotron emission \citep{jelic_et_al2008,bernardi_et_al2010}, all contaminants arising from foregrounds are thought to be spectrally smooth or faint enough not
to be problematic \citep{petrovic_oh2010}. The
brightness temperatures of these foregrounds can exceed the expected $\sim\!10~$mK amplitude of 
the 21cm EoR signal by up to five orders of magnitude 
\citep{zahn_et_al2007,santos_et_al2005}.  These foregrounds also serve as 
a source of noise that dominates the system temperature of 
radiometers in the $100-200~$MHz band.  

Projects aiming to detect the 21cm EoR signal have taken a variety of different approaches,
illustrating the breadth of parameter space available for designing such instruments.
Much of our discussion will focus on the PAPER experiment, but our results generalize to other arrays.
PAPER employs single-dipole antenna elements that are not steerable --- a design that
emphasizes spatially and spectrally smooth instrumental responses to facilitate
calibration and foreground removal \citep{parsons_et_al2010}.  This approach contrasts with the
large-dish approach taken by GMRT and the
station beam-forming approach taken by LOFAR and the MWA, where dipoles are added in-phase to
form more focused beams prior to correlation.
PAPER's design choice favors elements
with a single primary lobe of response horizon-to-horizon --- a choice which directly limits the collecting area
of each element.
On the one hand, for a fixed total collecting area, small antenna elements (such as PAPER's) increase the 
cost of correlation and imaging, owing to the $\mathcal{O}(N^2)$ scaling of the number of baselines
with number of antennas $N$.
On the other hand, it is imperative that imperfect calibration and sidelobes associated with 
station beam-forming 
(used by LOFAR and MWA)
not introduce spectral structure that
impedes foreground removal.
The trade-off of per-station collecting area versus primary-beam smoothness is one of the major
design parameters that must be addressed by first-generation experiments to lead into a
next-generation instrument such as phase-II of the Hydrogen Epoch of Reionization Array\footnote{http://reionization.org/} (HERA; \citealt{astro2010}) and the Square Kilometer Array\footnote{http://www.skatelescope.org/} (SKA).  
In all cases, first-generation experiments will be starved for sensitivity,
motivating the exploration of techniques for improving sensitivity.

Optimizing these design parameters for future 21cm EoR arrays requires a careful assessment of
the trade-offs between sensitivity and facility of calibration in first-generation experiments.  To
further this investigation, we provide in \S\ref{sec2} a detailed pedagogical derivation of the
sensitivity for an
interferometer targeting the 21cm reionization signal.  Within this section, we 
relate sensitivity to the number of repeated measurements of modes in the $uv$-plane, motivating 
sampling redundancy as an important metric of the sensitivity performance of an array.  
In \S\ref{sec:antenna_config}, we explore the conflicting goals of arrays aiming to characterize
foregrounds and those aiming to measure the evolving power spectrum of reionization fluctuations.
We then show how
antenna placement can be used to tune sensitivity relative to bright foregrounds.  We evaluate
several antenna configurations to arrive at a class of configurations that maximize sensitivity
to the 21cm EoR signal.
Finally, in \S\ref{sec:conclusion}, we discuss how our configuration studies,
along with experience with the technical 
challenges of foreground removal and correlating many antennas, will
influence the design of future experiments targeting the 21cm EoR signal.

\section{Sensitivity to the 21cm Power Spectrum}
\label{sec2}

Although derivations of the sensitivity of a radio interferometer to the expected
21cm EoR signal exist in the literature \citep{morales2005,mcquinn_et_al2006,pen_et_al2009},
we aim to clarify the derivation and be more precise about the approximations that 
have been made implicitly in previous derivations.
The goals of this section are to develop
a framework for comparing sensitivity to foregrounds that are often
related in Jy units, to highlight the effects of wide fields-of-view and wide bandwidths
on the approximations that are made, 
and to be as clear as possible about Fourier transform normalizations
while deriving the sensitivity of an
interferometric baseline to the three-dimensional power spectrum of 21cm reionization.

\subsection{Single-Baseline Response}
\label{sec:sense_baseline}

We begin with the basic definitions for the $3$D power spectrum of brightness temperature fluctuations 
(the statistic which 21cm efforts aim to measure)
and for the visibility (the fundamental observable of an interferometer).   
We then calculate the response of a single interferometric
baseline to the 21cm brightness-temperature fluctuations arising from cosmic reionization,
thereby deriving the relationship between visibilities calibrated to a ${\rm Jy}$ scale and reionization
fluctuations in $k$-space expressed in ${\rm mK}^2$ units.
For this derivation, we adopt a Fourier transform 
normalization convention that is consistent with that used in theoretical models 
(and is standard in cosmological work), but which 
differs from that used in radio astronomy.  With respect to the brightness temperature 
in a pixel of the sky plane/frequency data cube, $T(\vec x)$, and its Fourier dual $\tilde{T}(\vec k)$, this convention yields:
\begin{align}
\tilde{T}(\vec k)&=  \frac1{\mathbb{V}} \, \int T(\vec x) \, e^{-i\vec k\cdot\vec x}d^3 x\nonumber\\
T(\vec x) &=  \frac{\mathbb{V}}{(2 \pi)^3} \, \int \tilde{T}(\vec k) \, e^{i\vec k\cdot\vec x}d^3 k,
\end{align}
Here, $\mathbb{V}$ refers to the volume of the observed data cube and 
$\vec x$ is a $3$D vector that indicates direction on the sky 
and depth (the frequency dimension) within the field.  Likewise, $\vec k$ is a $3$D wave-vector
with projection $\vec k_\perp\equiv(k_x,k_y)$ in the plane of the sky, and $k_z$ along
the line-of-sight (frequency) direction.
We derive our response in the flat-sky approximation (as discussed below)
so that we may take $\vec x$ to be cartesian.

It follows from this convention that an estimate of the power spectrum is given by
\begin{equation}
    \widehat{P}(\vec k) \equiv \left\langle|\tilde{T}(\vec k)|^2\right\rangle  = \int{\widehat{\xi}(\vec r)\,e^{-i\vec k\cdot\vec r}d^3r},
    \label{eq:ft_cosmo}
\end{equation}
where angle brackets denote an ensemble average, $\vec r$ is the vector distance between two points, and $\widehat \xi(\vec r)$ is an estimate of the two-point correlation function of the 
measured $T$, given by
\begin{equation}
    \widehat{\xi}(\vec r)=\frac1{\mathbb{V}}\int{T(\vec x)T(\vec x + \vec r)d^3x}. \label{eqn:xi}
\end{equation}
It is important to note that the Fourier transform normalization defined by equation \ref{eq:ft_cosmo}
is {\it not} consistent with the transformation defining the
visibility for a single baseline \citep{morales_hewitt2004}; a volume factor, $\mathbb{V}$,
divides the integral in equation
\ref{eq:ft_cosmo}, but no observing volume appears in the denominator of
the following definition
of the visibility adapted from \citet{thompson1999}, their equation (2-21):
\begin{align}
    V(u,v,w,\nu)\equiv&\int{\frac{dl\ dm}{\sqrt{1-l^2-m^2}}A(l,m,\nu)I(l,m,\nu)}\nonumber\\
           &\times {e^{-2\pi i(ul+vm+w[\sqrt{1-l^2-m^2}-1])}},
    \label{eqn:vis}
\end{align}
where $\nu$ is spectral frequency, 
$l\equiv\sin\theta_x$ and $m\equiv\sin\theta_y$ are (in the small-angle approximation)
angular coordinates in image domain, $(u,v,w)\equiv\vec b/\lambda$ are the east-west, north-south, and
line-of-sight projections of baseline vector $\vec b$ toward a phase center, in units of 
observing wavelength $\lambda$,
$A(l,m,\nu)$ is a windowing function describing the field-of-view and bandpass response of an 
interferometer pair, and $I(l,m,\nu)$ is the specific intensity.

It is common to neglect the $(l,m)$-dependence of the $w$-term in the exponential ---
a simplification commonly referred to as the flat-sky approximation \citep{clark1999}.  This approximation is
valid within a radius $\approx\!10^\circ$ of the phase center or when phasing
to a direction orthogonal to the baseline vector (so that $w\approx0$).  For many low-frequency arrays,
including PAPER, wide fields-of-view make this approximation invalid, and 
proper imaging requires techniques such as
W-projection \citep{cornwell_et_al2003}.
However, the magnitude of the 21cm EoR power spectrum $P_{21}(\vec k)$ is not
expected to evolve significantly over $k\pm|k_\perp|$ for baselines shorter than 300 meters.
Hence, the linear combination of $\vec k$-modes 
generated by the point-spread function (PSF) of a baseline\footnote{
    From the perspective of deriving the power-spectrum response of a single baseline, the dominant
    effect of violating the flat-sky approximation is that the fringe-pattern of the baseline
    (which is a sinusoid in $l,m$) gradually de-tunes from a cartesian sinusoid away from phase center.
    As a result, a Fourier mode at $(u,v)$ that is sampled by a baseline will have a PSF
     in $\vec k$-space that
    is peaked in $\vec k_\perp$, but which
    includes contributions from modes with smaller $|\vec k_\perp|$ that project onto the fringe pattern
    nearer to the horizon.  The degree of peaked-ness depends on the relative gain of the primary beam
    within the region where the flat-sky approximation is valid.
    For nearly all of the $\vec k$-modes accessible to 21cm EoR instruments,
    $|\vec k|$ is dominated by the line-of-sight component $k_z$. It follows from
    equations \ref{eq:rcom_perp} and \ref{eq:rcom_los} that for line-of-sight scales arising from
    a 6-MHz bandwidth, the $\vec k_\perp$ component arising from a 300-m baseline 
    perturbs $|\vec k|$ by $\Delta\ln k\le0.5$, falling within the fiducial averaging interval
    used in \S\ref{sec:sense_independent_modes}.  This perturbation decreases rapidly for
    shorter baseline lengths.
}
is still representative of the statistical distribution described by $P_{21}(k)$.
We adopt the flat-sky approximation for simplicity, but use the full
area of the primary beam to estimate sensitivity.

To relate Fourier transform conventions that
differ by a factor of integration volume, 
we extend the definition of the 
visibility in the flat-sky approximation to include a similar Fourier transform along the frequency axis:
\begin{align}
    {\tilde V}(u,v,\eta)\approx&\int{dl~dm~d\nu~A(l,m,\nu)I(l,m,\nu)}\nonumber\\
        &\times {e^{-2\pi i(u \, l+v \, m+\eta \,\nu)}}.
    \label{eq:vhat_definition}
\end{align}
This definition ignores the frequency-dependence of $(u,v)$ arising from the changing
length of $\lambda$ dividing the physical separation of two antennas. 
21cm EoR experiments have 
large relative bandwidths, with $(u,v)$ varying by as much as 4\% over a 6-MHz bandwidth at $150~$MHz.  The 4\% variation in the $\vec k_\perp$ 
component of $\vec k$ that 
arises from the frequency-dependence of $(u,v)$ is smaller than the averaging interval used
later in \S\ref{sec:sense_independent_modes}, and approximation does not substantially
affect sampling of $P_{21}(\vec k)$, nor does it change the sensitivities we derive.
Of greater concern in our examination of the frequency-dependent sampling of the $uv$-plane by
a single baseline is the effect it may have on foreground removal.  We will revisit this issue
briefly in \S\ref{sec:antenna_config}, but we defer a detailed treatment of the subject to a future paper.

Squaring both sides and using $I=2k_{\rm B}T/\lambda^2$, with $\lambda$ being the mean wavelength
over the sub-band used in the Fourier transform, yields
\begin{align}
    {\tilde V}^2(u,v,\eta)&\approx \left(\frac{2k_{\rm B}}{\lambda^2}\right)^2
        \int{dl~dm~d\nu~dl^\prime~dm^\prime~d\nu^\prime}\nonumber\\
        &\times A(l,m,\nu)T(l,m,\nu)A(l^\prime,m^\prime,\nu^\prime)T(l^\prime,m^\prime,\nu^\prime)\nonumber\\
        &\times e^{-2\pi i[u \, (l-l^\prime)+v\,(m-m^\prime)+\eta \,(\nu-\nu^\prime)]}.
\end{align}
We now make the approximation that $A(l,m,\nu)$ is a top-hat windowing function.  
Explicitly integrating $A(l,m,\nu)$ determines the width and shape
of the convolution kernel in equation \ref{eq:v2_vs_pk_sinc}.
Since the width of this kernel is thereafter neglected and only enters later to tally the
number of independent wave-modes sampled, the top-hat approximation should be considered purely pedagogical.
Drawing $A(l,m,\nu)$ into the bounds of the integral yields:
\begin{align}
    {\tilde V}^2&(u,v,\eta)\approx\left(\frac{2k_{\rm B}}{\lambda^2}\right)^2
        \int_{(0,0,0)}^{(\theta,\theta,B)}\hspace{-0.5cm}dl~dm~d\nu
        \int_{(0,0,0)}^{(\theta,\theta,B)}\hspace{-0.5cm}dl^\prime~dm^\prime~d\nu^\prime\nonumber\\
        &\times T(l,m,\nu)T(l^\prime,m^\prime,\nu^\prime)e^{-2\pi i[u(l-l^\prime)+v(m-m^\prime)+\eta(\nu-\nu^\prime)]},
\end{align}
where $\theta\equiv\sqrt{\Omega}$, for primary beam field-of-view $\Omega$.
Changing variables so that $(l_r,m_r,\nu_r)=(l-l^\prime,m-m^\prime,\nu-\nu^\prime)$:
\begin{align}
    {\tilde V}^2&(u,v,\eta) \approx \left(\frac{2k_{\rm B}}{\lambda^2}\right)^2\bigg[
        \int_{(-\theta,-\theta,-B)}^{(0,0,0)}\hspace{-1.5cm}dl_r~dm_r~d\nu_r
        \int_{(0,0,0)}^{(\theta+l_r,\theta+m_r,B+\nu_r)}\hspace{-1.5cm}dl~dm~d\nu\nonumber\\
        &\hspace{1.2in}+\int_{(0,0,0)}^{(\theta,\theta,B)}\hspace{-1cm}dl_r~dm_r~d\nu_r
        \int_{(l_r,m_r,\nu_r)}^{(\theta,\theta,B)}\hspace{-1cm}dl~dm~d\nu\bigg]\nonumber\\
        &\times T(l,m,\nu)T(l\!-\!l_r,m\!-\!m_r,\nu\!-\!\nu_r)e^{-2\pi i(ul_r+vm_r+\eta\nu_r)}.
\end{align}
Integrating over $(l,m,\nu)$ and using equation \ref{eqn:xi} yields:
\begin{align}
    {\tilde V}_{21}^2(u,v,\eta)\approx&\left(\frac{2k_{\rm B}}{\lambda^2}\right)^2{\Omega\,B}
    \int_{(-\theta,-\theta,-B)}^{(\theta,\theta,B)}{\hspace{-0.5cm}dl_r~dm_r~d\nu_r}\nonumber\\
        &\times\widehat \xi_{21}(l_r,m_r,\nu_r)e^{-2\pi i(ul_r+vm_r+\eta\nu_r)},
\label{eq:fourier_transform}
\end{align}
where $B$ is the observing bandwidth, and 
where we now use the subscript ``21'' to make explicit that these quantities are derived for 21cm emission
from reionization.
Using $X$ and $Y$ to represent conversion factors from angle and frequency to comoving distance, respectively, we 
substitute $(Xl_r,Xm_r,Y\nu_r)$ for $(r_x,r_y,r_z)$ and $(Xk_x,Xk_y,Yk_z)$ for $2\pi(u,v,\eta)$.  
The factor of $2\pi$ follows from the cosmological Fourier convention.  This substitution then yields
\begin{equation}
    {\tilde V}_{21}^2(u,v,\eta)\approx\left(\frac{2k_{\rm B}}{\lambda^2}\right)^2\!\!\!\frac{\Omega\,B}{X^2Y}
    \!\int_{(-X\theta,-X\theta,-YB)}^{(X\theta,X\theta,YB)}{\hspace{-0.5cm}
    \widehat \xi_{21}(\vec r)e^{-i\vec k\cdot\vec r}d^3r},
\label{eq:fourier_transform_cosmo}
\end{equation}
This equation establishes the relationship between a $(u,v,\eta)$-mode measured by an interferometer and a $\vec k$-mode.
Hereafter we will use ``$(u,v,\eta)$-mode'' and ``$\vec k$-mode'' interchangeably to refer to a coherent
region in Fourier space.
Because the right-hand side of equation \ref{eq:fourier_transform_cosmo} 
is the Fourier transform of $\xi(\vec r)$ with a top-hat window, 
in Fourier space it becomes the convolution of the Fourier transform of these functions:
\begin{align}
    {\tilde V}_{21}^2(u,v,\eta)\approx&\left(\frac{2k_{\rm B}}{\lambda^2}\right)^2\!\!\frac{\Omega\,B}{X^2Y} 
    \Big[\widehat P_{21}(\vec k)*\big({\rm sinc}(2X\theta k_x)\nonumber\\
    &\times{\rm sinc}(2X\theta k_y)\,{\rm sinc}(2YBk_z)\big)\Big],
    \label{eq:v2_vs_pk_sinc}
\end{align}
where `$*$' signifies convolution in $\vec k$.
In the more general case this convolving kernel is not a sinc function, but 
the Fourier transform of the primary beam $\tilde A(l,m,\nu)$.
For primary beam responses larger than 30 arcmin,
the width of the kernel
in $k$-space 
is much smaller than the scales over which $P_{21}(\vec k)$ varies for 
the $k$-modes that are likely not to be dominated by foregrounds \citep{mcquinn_et_al2006}.  Thus, we drop the sinc kernel from equation \ref{eq:v2_vs_pk_sinc}, giving:
\begin{equation}
    {\tilde V}_{21}^2(u,v,\eta)\approx\left(\frac{2k_{\rm B}}{\lambda^2}\right)^2\frac{\Omega\,B}{X^2Y} \widehat P_{21}(\vec k).
    \label{eq:v2_vs_pk}
\end{equation}
Theoretical studies often express the $21$cm signal in a dimensionless manner given 
by $\widehat \Delta^2(k)\equiv \frac{k^3}{2\pi^2}\widehat P(\vec k)$ (using that 
$P_{21}({\vec k})$ is expected to be nearly isotropic; \citealt{mcquinn_et_al2006}), 
making it useful to write equation \ref{eq:v2_vs_pk} as
\begin{equation}
    {\tilde V}_{21}^2(u,v,\eta)\approx\left(\frac{2k_{\rm B}}{\lambda^2}\right)^2\frac{\Omega\,B}{X^2Y}\frac{2\pi^2}{k^3} \widehat \Delta_{21}^2(k).
    \label{eq:v2_vs_d2}
\end{equation}

\subsection{Single-Baseline Sensitivity Measuring One $\vec k$-mode}

The next step towards estimating the sensitivity to the 21cm signal is to calculate the power spectrum of the thermal noise of an instrument.  Thermal fluctuations produce a white-noise signal with root-mean-square (RMS) 
brightness temperature $T_{\rm N,rms}$, which in practice will be roughly equal to 
the sky temperature for 21cm instruments.  The thermal noise contributes a component to the
RMS amplitude of the visibility ${\tilde V}_{\rm N}$ equal to:
\begin{equation}
    {\tilde V}_{\rm N}=\frac{2k_B}{\lambda^2}T_{\rm N,rms} \, \Omega \, B.
    \label{eq:jy_temp_conv}
\end{equation}
This equation can be derived from equation \ref{eqn:vis} assuming a white-spectrum thermal noise for $I$ with temperature $T_{\rm N,rms}$.
We substitute ${\tilde V}_{\rm N}$ for ${\tilde V}$ in equation \ref{eq:v2_vs_d2} to get the 
noise contribution\footnote{
When squaring ${\tilde V}$ in equation \ref{eq:v2_vs_d2}, it is important to construct an estimator 
of $\Delta_{21}^2(k)$ that is not biased by
the noise power spectrum.  This can be accomplished by subtracting off a measured noise
power spectrum, or more elegantly by constructing cross-products ${\tilde V}_i{\tilde V}_j^*$ from
pairs of samples $i,j$ that measure the same Fourier mode but have independent thermal noise.
The sensitivities that are derived here reflect the residual error that remains in an \emph{unbiased}
construction of ${\tilde V}^2$.
} to the dimensionless power, $\Delta_{\rm N}^2(k)$, yielding:
\begin{equation}
    \Delta_{\rm N}^2(k)\approx X^2Y\frac{k^3}{2\pi^2}\Omega \,B  \,T_{\rm N,rms}^2(u,v,\eta).
    \label{eq:trms_vs_pk}
\end{equation}

Since there are $2Bt$ independent measurements of the noise for time $t$, the value of 
$T_{\rm N,rms}$ noise that should enter in equation \ref{eq:trms_vs_pk} is not the 
true temperature at any given time (which is usually called the system temperature $T_{\rm sys}$),  
but rather the error in how well $T_{\rm sys}$ can be measured (which relates to the error in how 
well thermal noise can be measured and subtracted off, and is $\sqrt{2}T_{\rm sys}$ for Gaussian
random noise) or 
$T_{\rm N,rms}^2=T_{\rm sys}^2/Bt$. 
With this substitution,
\begin{equation}
    \Delta_{\rm N}^2(k)\approx X^2Y\frac{k^3}{2\pi^2}\frac{\Omega}{2t}T_{\rm sys}^2,
    \label{eq:sensitivity}
\end{equation}
where $t$ is the integration time for sampling a particular $(u,v,\eta)$-mode, and
the factor of two in the denominator comes from the explicit inclusion of two
orthogonal polarizations to measure the total unpolarized signal.\footnote{
As defined above, $\Delta_{\rm N}^2(k)$ indicates the noise left in the map after one tries to 
subtract the noise power using all of the available information.  It may be defined equivalently as
relating to the signal-to-noise at which the true power spectrum, $\Delta_{21}^2(k)$, can be measured in a $k$-bin:
${\rm SNR}=\Delta_{21}^2(k)/\Delta_{\rm N}^2(k)$.
This definition assumes that $\Delta_{\rm N}^2$ is calculated for a real-valued sky, so that
baselines sampling positive and negative Fourier components are not counted as independent measurements.
} This equation differs from the derivations given in
\citet{morales2005} and \citet{mcquinn_et_al2006} by only this polarization factor.
Note how the 
power-spectrum sensitivity toward a particular $\vec k$-mode is independent of bandwidth, and that \citep{furlanetto_et_al2006}
\begin{align}
    X&\approx 1.9\left(\frac{1+z}{10}\right)^{0.2} h^{-1}\frac{\rm Mpc}{\rm arcmin}
    \label{eq:rcom_perp}\\
    Y&\approx 17\left(\frac{1+z}{10}\right)^\frac12
        \left(\frac{\Omega_m h^2}{0.15}\right)^{-\frac12}\frac{\rm Mpc}{\rm MHz}
    \label{eq:rcom_los},
\end{align}
giving us (for $\Omega_m=0.27$): 
\begin{equation}
    X^2Y\approx 540\left(\frac{1+z}{10}\right)^{0.9}\frac{h^{-3}\ {\rm Mpc}^3}{{\rm sr}\cdot{\rm Hz}}
    \label{eq:x2y_vs_z}.
\end{equation}
Substituting for $X^2Y$ at $z=8.5$ (assuming observations at 150 MHz) in equation 
\ref{eq:sensitivity}, and choosing fiducial PAPER parameters, we have:
\begin{align}
    \Delta_{\rm N}^2(k)\approx &~2.8\times10^4
    \left[\frac{k}{0.1h\ {\rm Mpc}^{-1}}\right]^3
    \left[\frac{\Omega}{0.76\ {\rm sr}}\right]^\frac32
    \nonumber\\&\times
    \left[\frac{T_{\rm sys}}{500\ {\rm K}}\right]^2
    \left[\frac{120\ {\rm days}}{t_{\rm days}}\right] 
    \left[\frac{|\vec u|}{20}\right]
    \ {\rm mK}^2,
    \label{eq:sensitivity_fiducial}
\end{align}
where we assume 120 days of observation with a baseline
of length $|\vec u|\sim20$ that allows $\sim\!13$ minutes of integration per day, for a total 
integration time of $9\times10^4$ seconds per $(u,v,\eta)$-mode.
In general, integration time per mode, per day depends strongly on baseline orientation and the latitude
at which an array is deployed.  We will estimate a minimum integration timescale here for
arrays at mid-latitudes, and defer an exact, configuration-dependent treatment until
\S\ref{sec:redundancy_sensitivity} and \S\ref{sec:antenna_config}.  We compute the amount of 
time a baseline samples 
a $(u,v,\eta)$-mode per day, $t_{\rm per\_mode}$, as it is limited by the timescale
for earth-rotation to move the sampling of a baseline a distance of $\Omega^{-1/2}$
in the $uv$-plane: $t_{\rm per\_mode}\approx1/\sqrt\Omega\omega_\oplus|\vec u|$, 
where $\omega_\oplus$ is the angular speed of the earth's rotation.
The choice of 20 wavelengths as a fiducial baseline length
is arbitrary, but represents an estimate of a minimum baseline length that is not dominated by
galactic synchrotron emission (see \S\ref{sec:maximum_redundancy}).

The cosmological 21cm signal is typically much smaller than the noise in a
single baseline, as given by equation \ref{eq:sensitivity_fiducial}.
This assumption is reflected in our derivation by the absence of sample variance as
a significant contribution to the errors we compute.
The globally averaged spin temperature of the 21cm transition is
$\langle T_{21}\rangle= 28~([1+z]/10)^{1/2}~{\rm mK}$ for neutral intergalactic medium,
assuming that the spin temperature of the 21cm transition is much
larger than the cosmic microwave background (CMB) temperature (which will almost certainly hold at
$z<10$; e.g. \citealt{furlanetto_et_al2006}).  For a patchy reionization process, an
estimate for the dimensionless power spectrum of 21cm fluctuations arising from 
inhomogeneities in the ionization fraction is given by
\begin{equation}
\Delta_{21cm}^2\sim\langle T_{21}\rangle^2 (x_H-x_H^2)/\ln(k_{\rm max}/k_{\rm min}),
\label{eq:model_eor_signal}
\end{equation}
where
$x_H$ is the average neutral hydrogen fraction and $k_{\rm min}$ and $k_{\rm
max}$ are the wave-vectors between which most of the power lies.  Consistent
with this estimate, when $x_H\approx0.5$, simulations of reionization find 
$10<\Delta_{21}^2<100~{\rm mK}^2$ with a flat spectrum over 
$0.1< k<10~h\ {\rm Mpc}^{-1}$ \citep{mcquinn07,trac07}. 
Models with rarer sources tend to produce larger ionized
regions and more power than those with more abundant sources \citep{mcquinn07}.
Comparing equation \ref{eq:model_eor_signal} to the sensitivity of a baseline
in equation \ref{eq:sensitivity_fiducial} motivates the exploration of
methods for bolstering the sensitivity of instruments to the 21cm EoR
signal.

Before proceeding, it is worth reiterating the assumptions that went into the previous derivation,
and to consider any generality that may have been lost:  
\begin{enumerate}
\item We assumed we could work in the 
flat sky limit, which we justified by noting that the 21cm EoR power spectrum $P_{21}(k)$ is not
expected to evolve on the scale of the mode-mixing introduced by this approximation.  
\item We
ignored the frequency dependence of the $(u,v)$-coordinates of a baseline for the same reason.  For
baselines longer than $\approx 300$m, both of the above assumptions break down and cause errors at
the several percent level.  
\item We pedagogically treated the antenna primary beam as top-hat
function, but argued that any primary beam with FWHM wider than 30 arcmin creates a sufficiently small
convolving kernel in equation \ref{eq:v2_vs_pk_sinc} that its shape may be neglected.  
\item We have assumed the SNR in any individual $\vec k$-mode measurement to be much less than unity.
Since any additional improvement to sensitivity comes from independent $\vec k$-modes whose inclusion
beats down both thermal fluctuations and sample variance, this assumption in effect
allows us to ignore sample variance as a significant source of error.
\item Finally, our
value of observing time per mode in equation \ref{eq:sensitivity_fiducial} represents a lower bound;
its exact value generally depends on baseline orientation and array latitude and must be computed
explicitly for specific EoR experiment locations and configurations.
\end{enumerate}

\subsection{Combining Independent $\vec k$-mode Measurements}
\label{sec:sense_independent_modes}

With the sensitivity of one baseline to one $\vec k$-mode derived, we now turn our attention
to the sensitivity boost that comes from combining multiple baselines.  In this section, we
consider an analytically tractable case, where each baseline measures an independent
$\vec k$-mode.  Statistically independent $\vec k$-modes can be combined to improve sensitivity
proportionally to the square root of the number of samples, $N_s^{1/2}$.
We ignore sampling redundancy --- the possibility that many baselines can measure the same 
$\vec k$-mode --- which closely approximates the response of minimum-redundancy
arrays used for imaging (see \S\ref{sec:minimum_redundancy}).  
Although somewhat contrived, this example demonstrates how several different sensitivity boosts
that arise from system and observing parameters.
In \S\ref{sec:redundancy_sensitivity} we will 
use numerical simulations to calculate the sensitivities of real array configurations, including 
sampling redundancy.

Several assumptions are used to make this derivation tractable. 
The final expression derived in this section ---
equation \ref{eq:multimode_sensitivity_fiducial} --- is not intended to be generally applicable, but rather to 
illustrate the different effects that come into play when combining measurements.  The fully general 
case is presented in equation \ref{eq:multimode_redundant_sensitivity_fiducial}, where one must 
numerically calculate the effects of array configuration.  Our principal assumption is that our 
baselines uniformly sample the $uv$-plane within a radius $u_{\rm max}$.  As before, we also assume 
that our baselines are short enough to neglect to contribution of $k_{\perp}$ to $\vec k$, 
generally true for baselines under 300m.  Finally, using PAPER as a model, we assume an array at 
$45^\circ N/S$ latitude
observing for six sidereal hours per day ($t_{\rm per\_day}$ below) during which colder patches of 
the synchrotron sky are overhead.  Since we assume no sampling redundancy, it is irrelevant whether
these six hours of observation are spent tracking a single phase center or are broken
up into several observations with different pointings.

Before we discuss the different sources of independent $\vec k$-mode samples, we define a fiducial
measurement which all improvements are relative to.  In this section, we use equation \ref{eq:sensitivity_fiducial}, the sensitivity of one baseline measuring one $\vec k$-mode as our benchmark.  We will refer to the new noise level after combining measurements as $\Delta_{\rm N}^2(k)$ and express this value relative to our fiducial value, $\Delta_{\rm{N},0}^2(k)$.

We now will outline the different sources of independent $\vec k$-mode samples and present physical
arguments for their dependencies on various parameters.  A full derivation, including the prefactor of the 
final sensitivity
result in equation \ref{eq:multimode_sensitivity_fiducial}, is presented in Appendix 
\ref{app:multimode_sensitivity}.  

\begin{enumerate}

\item \textbf{Multiple line-of-sight samples} One source of independent $\vec k$-mode samples comes from the many line-of-sight Fourier modes
measured by a single baseline; since $\Delta_{21}^2(k)$ is expected to evolve on log-$k$ scales,
data may be binned in equally-spaced $\Delta{\rm ln}~k$ intervals, so that $N_s(k)\propto k$.
For example, with 6-MHz observing bandwidth, $B$, a single baseline will
sample $k\approx0.06h\ {\rm Mpc}^{-1}$ once, $k\approx0.12h\ {\rm Mpc}^{-1}$ twice, etc.
This linearly
increasing number of independent samples versus $k$ produces a ${\rm SNR}\propto k^{1/2}$ scaling.
The number of samples within a bin is also dependent on the bin size, giving rise to a final
proportionality after combining line-of-sight modes:
\begin{align}
\Delta_{\rm N}^2(k) \propto \left[\frac{1}{k}\right]^\frac12 \left[\frac{1}{B}\right]^\frac12 \left[\frac{1}{\Delta\ln k}\right]^\frac12 \Delta_{\rm{N},0}^2(k)
\end{align}

\item \textbf{Multiple time samples} Another source of independent measurements comes from 
the number of time bins available for measuring $\Delta_{21}^2(k)$ in a sidereal day.  These
additional samples grow linearly with the daily observation length, $t_{\rm per\_day}$.  (Accumulating
samples over multiple days was already accounted for in equation \ref{eq:sensitivity_fiducial}.)
Therefore, the sensitivity increases as:
\begin{align}
\Delta_{\rm N}^2(k) \propto \left[\frac{1}{t_{\rm per\_day}}\right]^\frac12 \Delta_{\rm{N},0}^2(k)
\end{align}

\item \textbf{Multiple uv-plane samples} A final source
of independent samples comes from baselines sampling different regions of the $uv$-plane.
The most straightforward way these samples affect the sensitivity is through adding more baselines.
Each baseline is an independent measurement, so sensitivity grows as the square root of number of
baselines, or, linearly with the number of antennas, $N$.

Secondly, we need to add up all the measurements across the $uv$-plane.
Using our assumption that $uv$-samples are uniformly distributed within a
circle of radius $u_{\rm max}$ in the $uv$-plane,
we integrate contributions from rings of constant $|\vec u|$ up to
a distance $|\vec u|=u_{\rm max}$. 
This integration is simplified by noting that 
each ring of constant $|\vec u|$ contributes equally to the sensitivity of the final
measurement; as $|\vec u|$ increases, the reduction in coherent integration time per $(u,v,\eta)$-mode 
is offset by the increasing number of baselines sampling within that ring.  
Integrating a constant sensitivity contribution versus $|\vec u|$ gives rise to a
$u_{\rm max}^{1/2}$ term in
the resulting residual noise estimate for 
minimum-redundancy arrays.

There is also
a factor of solid beam angle $\Omega^{-1/4}$, which is a combination of two factors.
First, there is a decrease
in integration time per $(u,v,\eta)$-mode associated with a broader primary beam, which
scales as the width of the primary beam, $\Omega^{-1/2}$.
This term is somewhat offset by a second factor: the increased number of independent modes sampled, 
contributing a
factor of $\Omega^{1/4}$ to sensitivity.
The result is:
\begin{align}
\Delta_{\rm N}^2(k) \propto u_{\rm max}^\frac12 \left[\frac{1}{\Omega}\right]^\frac14 \left[\frac{1}{N}\right]^\frac12 \Delta_{\rm{N},0}^2(k)
\end{align}

\end{enumerate}

Combining all these different gains from binning the data with the prefactor calculated in 
Appendix \ref{app:multimode_sensitivity} yields the final result of this section:
\begin{align}
    \Delta_{\rm N}^2(k)\approx &~150   
    \left[\frac{k}{0.1h\ {\rm Mpc}^{-1}}\right]^\frac52
    \left[\frac{6\ {\rm MHz}}{B}\right]^\frac12
    \left[\frac1{\Delta{\rm ln}~k}\right]^\frac12
    \nonumber\\&\times
    \left[\frac{\Omega}{0.76\ {\rm sr}}\right]^\frac54
    \left[\frac{T_{\rm sys}}{500\ {\rm K}}\right]^2
    \left[\frac{6\ {\rm hrs}}{t_{\rm per\_day}}\right]^\frac12
    \nonumber\\&\times
    \left[\frac{120\ {\rm days}}{t_{\rm days}}\right]
    \left[\frac{u_{\rm max}}{20}\right]^{\frac12}
    \left[\frac{32}{N}\right]
    \ {\rm mK}^2,
    \label{eq:multimode_sensitivity_fiducial}
\end{align}

\subsection{The Sensitivity Benefits of Redundant $uv$-Sampling}
\label{sec:redundancy_sensitivity}

As mentioned above, sensitivity to $\Delta_{21}^2(k)$ depends on both the sensitivity obtained 
in individual $(u,v,\eta)$-mode bins and 
the number of bins sampled. As discussed in the context of CMB
analysis, sensitivity is most efficiently improved by integrating coherently on select
modes until ${\rm SNR}\sim1$ is obtained, whereupon sampling additional modes 
to beat down sample variance in the cosmological signal becomes the most efficient
way of improving sensitivity \citep{park_et_al2003}.  Equation \ref{eq:sensitivity_fiducial} suggests
that the PAPER experiment, along with many other first-generation 21cm EoR experiments \citep{mcquinn_et_al2006},
will be firmly in the ${\rm SNR}<1$ regime for individual modes for the near future.  
As a result, it is natural to explore how sensitivity
might be improved by choosing antenna configurations that maximize the degree to which
$uv$-bins are sampled by multiple baselines.

In this section, we outline a formalism for computing the sensitivity boost of a generic 
antenna array, expressed in terms of a redundancy metric that tracks sensitivity
relative to a fiducial measurement, which 
we choose to be a single baseline with a one-second integration.
The choice of fiducial integration time is arbitrary, but affects the scaling constants in
the equations that follow.


Next, we define a metric for the sampling redundancy generated by an array,
\begin{equation}
    \frac{f}{f_0}\equiv\frac{\sum_i{n_i^2}}{\sum_i{n_i}}.
\label{eq:redundancy_metric}
\end{equation}
where $f_0$ is the sampling redundancy of a single baseline with a one-second integration
and $n_i$ represents the number of one-second samples falling in $uv$-bin $i$.  The ratio $f/f_0$
measures the increase in sensitivity for a redundant array over one in which there is no
sampling redundancy.  We motivate the definition of sampling redundancy in Appendix 
\ref{app:multimode_redundant_sensitivity} and show that sensitivity increases as $\sqrt{f/f_0}$.  

The instantaneous (single-integration) redundancy of an array
ranges from $f_0$ to $Nf_0$, where 
$N$ is the number of antennas in
an array configured for the densest non-overlapping packing of antennas possible in two dimensions: a 
filled circular aperture; we derive this case in Appendix \ref{app:circ_ap_deriv}.
However, computing the instantaneous redundancy of an array 
does not account for any additional redundancy that may be
generated through earth-rotation synthesis. A baseline sampling a non-redundant $uv$-bin can, some time
later, migrate in the $uv$-plane to sample a bin already sampled by another baseline.  
Generally, the redundancy generated through earth-rotation synthesis depends strongly on antenna configuration.
We will rely on
numerically computed redundancies for specific configurations.  

We include the effect of sampling redundancy by using $f$ defined for
one-second integrations toward a transiting phase center, calculated from equation
\ref{eq:sensitivity}.  Our result in equation \ref{eq:multimode_redundant_sensitivity_fiducial}
(which is derived in full in Appendix \ref{app:multimode_redundant_sensitivity}), 
is expressed in terms of fiducial observation
and array parameters.
 Unlike equation \ref{eq:multimode_sensitivity_fiducial}, 
the effects of array configuration are now captured in the computed value for $f$:
\begin{align}
    \Delta_{\rm N}^2(k)\approx &~60
    \left[\frac{k}{0.1h\ {\rm Mpc}^{-1}}\right]^\frac52
    \left[\frac{6\ {\rm MHz}}{B}\right]^\frac12
    \left[\frac1{\Delta{\rm ln}~k}\right]^\frac12
    \nonumber\\&\times
    \left[\frac{\Omega}{0.76\ {\rm str}}\right]
    \left[\frac{T_{\rm sys}}{500\ {\rm K}}\right]^2
    \left[\frac{6\ {\rm hrs}}{t_{\rm per\_day}}\right]^\frac12
    \nonumber\\&\times
    \left[\frac{120\ {\rm days}}{t}\right]
    \left[\frac{32}{N}\right]
    \left[\frac{10^4 f_0}{f}\right]^\frac12
    \ {\rm mK}^2,
    \label{eq:multimode_redundant_sensitivity_fiducial}
\end{align}
The value of $f/f_0$ varies substantially with $N$ and for different 
antenna configurations.  An array without redundant sampling will have $f/f_0 = 1$.
A nominal value of $f/f_0=3.4\times10^4$, representative of the
antenna configurations considered later,
yields $\Delta_{\rm N}^2(k)\approx33 {\rm mK}^2$ at $k=0.1h {\rm Mpc}^{-1}$.
As described in \S\ref{sec:sense_independent_modes}, we assume six-hour observations.
For PAPER, these observations are phased to transit pointings separated by two hours and
are accumulated into separate $(u,v,\eta)$-bins for each pointing.  (Two hours corresponds to the
approximate width of the PAPER primary beam, after which a new, statistically independent 
region of sky dominates the data).  Since there can be no redundancy between samples from different pointings,
this has the effect of somewhat reducing $f$.  Generally, $f$ accounts for most
effects relating to observing strategy.

\section{Antenna Configuration Studies}
\label{sec:antenna_config}

Designers of interferometric arrays for sonar, radar, and radio astronomy applications 
have long appreciated the necessity of carefully choosing the physical placement of
array elements to produce desirable samplings of the $uv$-plane.  One of the most
popular criteria --- the minimization of image-domain sidelobes arising
from incomplete sampling --- motivates array designs that maximize the number of independent Fourier modes sampled,
or equivalently, minimize the redundancy with which $uv$-pixels are sampled.  Such
{\it minimum-redundancy} configurations are valuable for characterizing
point-source foregrounds to the 21cm EoR signal, since each $uv$-pixel provides
unique information for constraining the image-domain distribution of flux density.  
All sampled Fourier modes contribute to each
image-domain location, making sensitivity
independent of antenna arrangement within a fixed maximum baseline length for image-domain measurements.  
This gives rise to the traditional adaptation of the
radiometer equation for interferometers (see \citealt{wrobel_walker1999}, their equation 9-23):
\begin{equation}
    T_{\rm rms}\Omega_{\rm s}=\frac{T_{\rm sys}\Omega}{\sqrt{BtN(N-1)}}
    \label{eq:interferometric_radiometer}
\end{equation}
where $\Omega_{\rm s}$, the angular size of a synthesized beam, is implicitly related to
maximum baseline length.

In contrast, the sensitivity of Fourier-domain measurements {\it do} depend dramatically
on array configuration.  First-generation
experiments will constrain the power-spectrum of 21cm EoR
fluctuations by sampling $\vec k$-modes accessed
via spectral structure in sampled $uv$-pixels.  As discussed in
\S\ref{sec:redundancy_sensitivity}, sensitivity-limited arrays will do best by
redundantly sampling a select number of $uv$-pixels.  However, {\it maximum-redundancy} array
configurations run directly counter to
the needs of image-domain work, and will look counter-intuitive to those familiar
with standard minimum-redundancy array configurations.

In \S\ref{sec:sense_baseline}, we discussed
how a single baseline measures $\Delta_{21}^2(k)$ at a range of $k$-scales with the
approximation that $(u,v)$ are not frequency dependent, and argue that this approximation does
not dramatically affect response to the 21cm EoR signal.  The impact of frequency-dependent
sampling on foreground response is somewhat more concerning, and has been used to
argue for configurations that 
produce uniform sampling of the $uv$-plane \citep{bowman_et_al2009}.  Such sampling could permit
chosen $uv$-modes to be sampled continuously versus frequency, even if the baseline
sampling them changes.
In a future
paper, we will explore in detail the effects of frequency-dependent $uv$-sampling,
showing that for baselines shorter than $\sim\!100$ wavelengths, all but the smallest $k$-modes
are accessible using the inherent frequency-dependent $uv$-sampling produced by a baseline.
This forthcoming result contrasts with the view that 21cm EoR arrays
must produce uniform $uv$-coverage and motivates the exploration of other maximum-redundancy
array configurations.
With an eye toward using the inherent frequency-dependent sampling of each baseline independently to sample 
$\Delta_{21}^2(k)$, we largely ignore the frequency dependence of
array sampling in the discussion of maximum-redundancy arrays; redundant sampling
will be redundant at all frequencies. 

Both minimum- and maximum-redundancy configurations have valuable properties
for 21cm EoR work.  Array configurations aiming to incorporate aspects
of both must attempt to strike a balance between their opposing influences.
Where this balance lies depends on the relative immediacy of sensitivity and
foreground-removal needs.  Given our current ignorance of many foreground
properties, it is most straightforward to consider each type of configuration
separately, as we will below.

\subsection{Minimum-Redundancy Array Configurations}
\label{sec:minimum_redundancy}

Designing a minimum-redundancy antenna configuration reduces to choosing a
real-valued sampling function $\mathcal{A}(x,y)$, with Fourier dual 
$\tilde{\mathcal{A}}(u,v)$, such that $\int{\big|\tilde{\mathcal{A}}\big|^2du\
dv}$ is minimized.  This optimization problem is usually discretized by
sampling the aperture plane on the scale of the aperture of a single antenna
element, and by assuming $\mathcal{A}(x,y)$ to be unity-valued at a location containing
an antenna element and zero-valued elsewhere.  We are often
interested in dense packings of antennas that also minimize the maximum distance
of $uv$-samples from the origin.
Compact packings of antennas have the desirable property of sampling nearly all
Fourier modes for a targeted angular resolution.  Compact
minimum-redundancy configurations can also be trivially scaled to larger
physical spacings to sample smaller angular scales.

\begin{figure*}\centering
    \includegraphics[width=0.8\columnwidth]{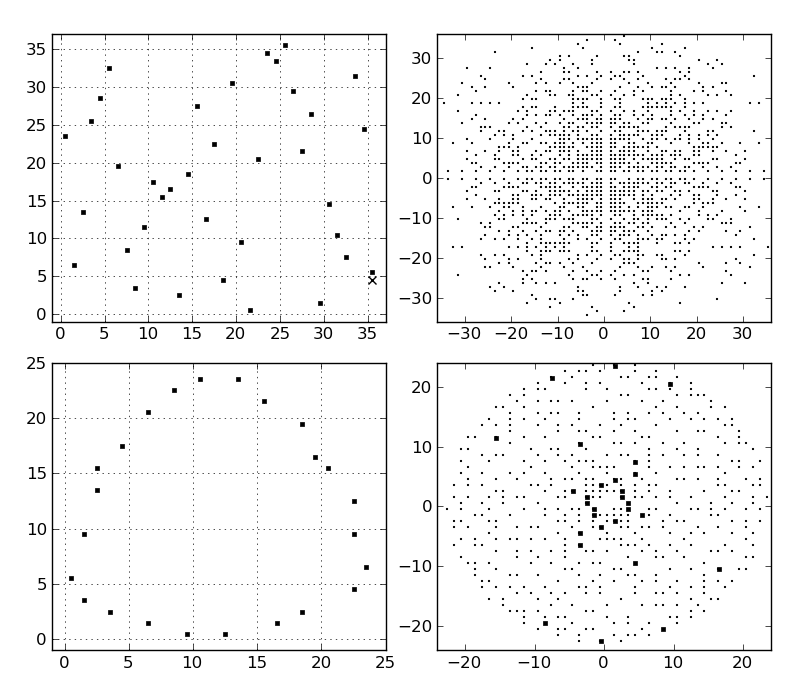}\caption{
The top-left panel shows a minimum-redundancy antenna configuration based on 
an $N=36$ Costas array (see \S\ref{sec:minimum_redundancy}).
The associated instantaneous zenith-phased coverage of the $uv$-plane is shown in the top-right panel.
For comparison, the bottom-left panel shows a 24-antenna configuration derived from a 
Reuleaux triangle that was iteratively optimized to generate uniform
$uv$-coverage \citep{keto1997}.  This configuration has been scaled for comparison to the size of a $24\times24$
Costas array to remove the effect of physical scale on the redundancy metric.  The $uv$-coverage
of this configuration (bottom-right panel) highlights the 28 redundant samplings with larger dots.
Contrast this with Costas arrays, which are perfectly non-redundant 
and have simple construction algorithms for numbers of
antennas near prime numbers.  The ``x'' in the top-left panel shows where a 37th antenna
can be introduced without generating redundant sampling if we relax the
constraint that Costas arrays must have only one antenna per row and column.  The antenna layout
in the top-left panel, including the additional antenna,
is scaled to grid spacings of 3.75 meters and 8 meters in the configurations labeled 
\texttt{min37c} and \texttt{min37},
respectively, in Figures \ref{fig:config_study_configs}, \ref{fig:config_study_uv}, 
\ref{fig:rprofile}, and \ref{fig:config_study_sense}.
    }\label{fig:costas_minred}
\end{figure*}

A fact that may be under-appreciated in the radio astronomy community is that
this optimization problem has may parallels with Golomb rulers
\citep{sidon1932,babcock1953}, Golomb rectangles \citep{robinson1985}, and
Costas arrays \citep{costas1984} --- mathematical constructions
originally motivated by radar and sonar applications.  Investigation of Golomb
rulers and Costas arrays are active fields of mathematical research with
interesting applications \citep{golomb_gong2004}.  In particular, the study of
Costas arrays ($N\times N$ matrices with elements chosen such that no two
elements share a row or column and such that the displacement vector between each pair
of elements is unique) has yielded algorithms for generating minimum-redundancy arrays where
$N$ is near a prime number \citep{golomb_taylor1984}.  For generating
array configurations, directly computing antenna locations following construction algorithms
for Costas arrays is a vast improvement over the iterative optimization approaches presented in the literature
\citep{keto1997,devilliers2007}.

Although Costas arrays do not quite capture the full minimum-redundancy
optimization problem (they omit samplings along the $u$ and $v$ axes and they 
do not attempt to optimize how compactly antennas are placed), they
do sample approximately one-quarter of the available $uv$-plane without redundancy.  
This filling fraction exceeds what has been demonstrated with other approaches in the literature.
For comparison, we examine the dithered Reuleaux-triangle
approach favored by \citet{keto1997} for generating configurations with uniform $uv$-coverage,
scaled to the size of an equivalent Costas array (in this case, for $N=24$) to
remove scale-dependence in the redundancy metric.  As we show in Figure \ref{fig:costas_minred}, this configuration
redundantly samples 28 locations with its instantaneous zenith-phased $uv$-coverage.  For imaging
point sources in the high-SNR limit,
this configuration loses 6.7\% of the information accessed by a roughly equivalent
configuration derived from a Costas array.

As an example of a larger-sized minimum-redundancy configuration derived from a Costas array,
we produce the $N=36$ antenna configuration shown in the upper-left panel of Figure \ref{fig:costas_minred}
following the Welch construction \citep{golomb_taylor1984},
where $N$ is chosen to be one less than the prime $p=37$.
According to this construction, we choose an integer $\alpha=35$ with the property that
$0<\alpha\leq p$ such that
$\alpha^N{\rm mod}\ p=1$ and $\alpha^i{\rm mod}\ p\ne1$ for $0<i<N$.
This construction produces row and column indices $(i,\alpha^i{\rm mod}\ p)$
for placing antennas on an $N\times N$ grid.  Note that
$(i,\alpha^{i+j}{\rm mod}\ p)$ also produces a Costas array for $0\le j<p$. Figure \ref{fig:costas_minred}
illustrates the antenna configuration generated from a Costas array with $j=23$, chosen so that
the resulting configuration could be augmented with one more antenna (see the upper-left
panel of Figure \ref{fig:costas_minred}) without incurring
any redundancy\footnote{
    For certain Costas arrays, relaxing the restriction that no two elements share a row or column
    allows one more antenna to be placed within the $N\times N$ matrix of possible 
    locations, generating new $uv$-samples without incurring sampling redundancy.  
    Such augmentations can be tested for in a Costas array by first
    computing the $uv$-sampling matrix for a Costas array (done by convolving the antenna placement
    pattern with itself) and then convolving the result with the original antenna placement
    pattern.  A zero value within the original $N\times N$ matrix indicates
    a location where an antenna can be added without increasing sampling redundancy.
    Many Costas arrays may be created to test for augment-ability by
    trying all valid $\alpha$ and $j$, as defined above.
}.
For comparison with other antenna configurations, we generate two realizations of this 
minimum-redundancy configuration: one using
3.75-meter spacing between rows and columns (\texttt{min37c}), and one
using 8-meter spacing that would be of more practical use for imaging
foregrounds (\texttt{min37}).  The performance of these configurations are compared with maximum-redundancy
configurations
in Figures \ref{fig:rprofile} and \ref{fig:config_study_sense}.  Costas arrays efficiently
generate minimum-redundancy arrays for imaging, but these figures demonstrate
that the minimal redundancy of these arrays has negative repercussions for power-spectrum sensitivity.

\subsection{Maximum Redundancy Arrays}
\label{sec:maximum_redundancy}

As shown in \S\ref{sec:redundancy_sensitivity}, sensitivity may be gained by focusing
limited collecting area on specific modes of the power spectrum.
However, the rising contribution of galactic synchrotron emission at large angular
scales, the dominance of point-source emission at small angular scales, and the
expectation that low-order line-of-sight (smooth-frequency) components must be
used to suppress foregrounds suggest that array configuration must be
informed by foreground characterization.
Compact antenna configurations improve sensitivity by increasing sampling redundancy, 
motivating the centrally-condensed configurations explored for
interferometers targeting the 21cm EoR signal \citep{bowman_et_al2006,lidz_et_al2008}.  Such configurations,
however, suffer from a number of practical deficiencies for this application:
\vspace{-4pt}
\begin{enumerate}
\itemsep 0pt
\parskip 0pt
\item The low fringe-rates associated with baselines sampling small $|\vec u|$ modes impede the discrimination
of celestial signals from instrumental systematics (e.g. crosstalk).
\item Phase and gain self-calibration are compromised by the lack of longer baselines.
\item Proximity of antenna elements can cause cross-coupling, producing antenna-specific 
deviations in primary beam response.
\item Foreground emission peaks in brightness at the small $|\vec u|$-modes that are most
heavily sampled by centrally-condensed configurations.
\end{enumerate}
\vspace{-4pt}

Phase switching can help mitigate (though not eliminate) crosstalk, and incorporating a
relatively small number of antennas at longer spacings can improve phase and gain calibration from point sources.
Increasing the spacing between densely packed antennas can substantially decrease cross-coupling at 
the expense of the 
redundancy generated from earth-rotation synthesis.  The fact that foreground
emission peaks at small $|\vec u|$ is unavoidable.

\begin{figure*}\centering
    \includegraphics[width=0.8\columnwidth]{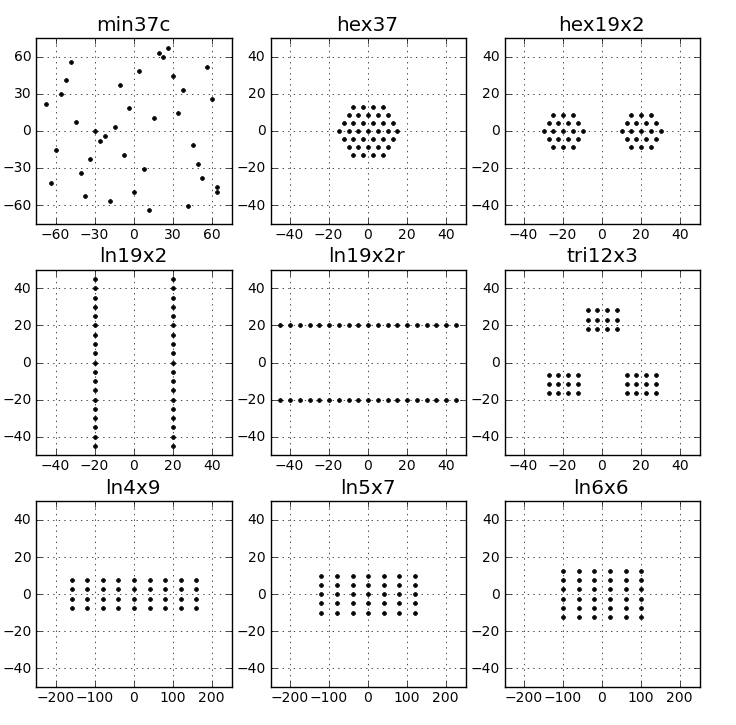}\caption{
Shown above are the north-south (vertical axis) and east-west (horizontal axis) antenna positions
in meters for various fiducial array configurations. 
With the exception of \texttt{min37c},
these arrangements aim to improve power-spectrum
sensitivity in the regime where errors are not dominated by sample variance in the
cosmological signal by redundantly sampling regions in the $uv$-plane with many baselines.  
The different
configurations explore different design ideas, including how
maximum-redundancy arrays may be generated for regions farther from the center
of the $uv$-plane.  In contrast, \texttt{min37c} (see Figure \ref{fig:costas_minred})
is an array configuration tuned to
minimize sampling redundancy, thereby 
improving imaging
by maximizing the number of independent measurements of the $uv$-plane (see \S\ref{sec:minimum_redundancy}).
The $uv$-sampling patterns generated by these configurations are shown in
Figure \ref{fig:config_study_uv}.  
    }\label{fig:config_study_configs}
\end{figure*}

\begin{figure*}\centering
    \includegraphics[width=0.8\columnwidth]{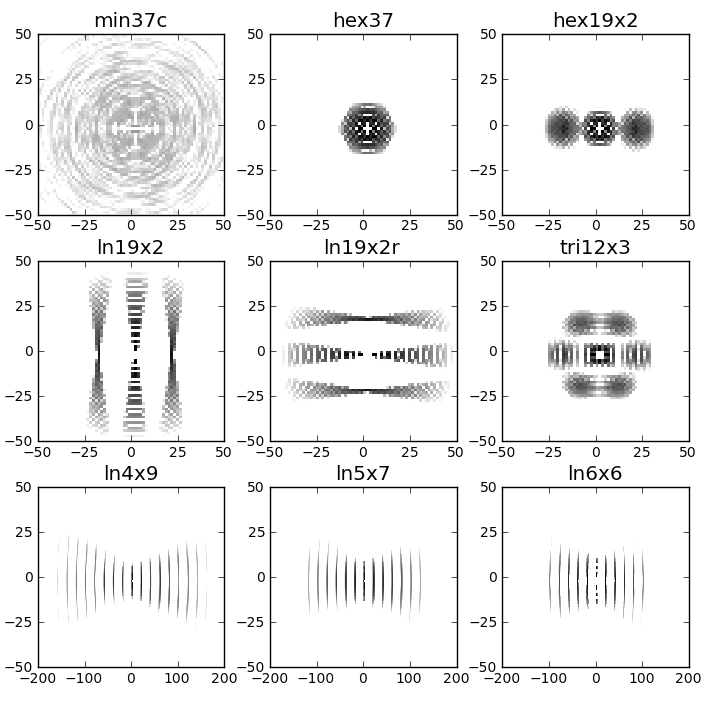}\caption{
The $uv$-plane samplings shown above were generated for each of the antenna configurations
from Figure \ref{fig:config_study_configs}, assuming a 2-hour observation at $40^\circ N$ latitude
of a zenith-transiting phase center at 150 MHz.  Sampling is
plotted using $uv$-bins that are 1.5 wavelengths on a side, with color scale indicating ${\rm log}_{10}$
of the number of one-second samples falling in each bin, ranging from 3 (white) to 5 (black). 
Antenna configurations generating redundant $uv$-sampling patterns increase
sensitivity to particular Fourier modes used to probe the $\Delta_{21}^2(k)$ power spectrum 
at the expense of sampling
multiple modes.  
Several of the 
configurations illustrated above direct sensitivity toward regions at higher $|\vec u|$, thereby avoiding
the instrumental systematics and brighter foregrounds associated with sampling near the origin of the $uv$-plane.
The redundancy metrics (see \S\ref{sec:redundancy_sensitivity}) computed for these sampling patterns are
shown in Figure \ref{fig:rprofile}.
    }\label{fig:config_study_uv}
\end{figure*}

Fortunately, for measuring a physical scale at reionization, a 21cm
EoR experiment has considerable flexibility in choosing an angular scale, making it
possible to generate antenna configurations that reap many of the sensitivity benefits of
centrally-condensed configurations but which avoid some of the associated deficiencies by directing sensitivity
toward higher-$|\vec u|$ modes.  Consider, for example, a configuration consisting of two clusters 
of $N/2$ closely-packed antennas
whose centers are separated by a distance larger than the diameter of each cluster 
(see \texttt{hex19x2} in Figure \ref{fig:config_study_configs}).  Excluding the central region of 
the $uv$-plane sampled by intra-cluster antenna pairs, we see in Figure \ref{fig:config_study_uv} 
that $N^2/4$ samples are concentrated in
a region near $|\vec u|=30$.  In this region, excluding earth-rotation synthesis, we compute from equation
\ref{eq:f_disk} that $f/f_0=N/2$.  Including redundancy generated by the earth's 
rotation over the course of a two-hour observation toward a transiting phase center, we compute 
$f/f_0=1.1\times10^4$ following equation \ref{eq:redundancy_metric}.
Furthermore, by adjusting the spacing between antenna clusters, this configuration 
can be tuned to focus sensitivity to regions of the $uv$-plane where 
galactic synchrotron and point-source foreground emission are minimized.

The \texttt{hex19x2} design described above and shown in Figure
\ref{fig:config_study_configs} can be improved upon in several ways.  Firstly,
perturbing the shape of antenna clusters can improve overlap resulting
from earth-rotation synthesis.  Long rows of antennas (see \texttt{ln19x2} in Figure
\ref{fig:config_study_configs}) do well for this; as the
Earth rotates, this sampling pattern slides over itself along the longest
axis.  As a result, the new $uv$-samples generated are largely
redundant with regions that have already been previously sampled (see Figure \ref{fig:config_study_uv}).
It should be noted that the latitude at which an array is deployed influences 
the design of maximum-redundancy array configurations.
At latitudes near $45^\circ N/S$, the performance of the row-based
configurations we explore
is largely independent of the orientation of the rows.  For arrays near the equator,
rows oriented east-west yield better sensitivity because the spacing between rows
is maintained through earth-rotation synthesis.  In these cases, however, the slow fringe-rates
of the north-south baselines generated may make them more susceptible to instrumental systematics.
Motivated by the locations of current 21cm EoR arrays, we have restricted ourselves 
to considering only mid-latitudes. 

An additional
optimization relates to exploiting the Hermitian-symmetry of the $uv$-plane for a real-valued sky:
a baseline vector $\vec u$ also samples the $uv$-plane at $-\vec u$.  As a result, for sampling
the $uv$-plane at the location of the displacement between rows, the interior
rows in antenna arrangements such as \texttt{ln4x9}, \texttt{ln5x7}, and \texttt{ln6x6} in Figure
\ref{fig:config_study_configs} are used in two pairings --- once with each adjacent row.  For 
$N$ antennas arranged into $R$ rows,
$R-2$ rows are used in both positive and negative pairings,
generating a peak instantaneous 
redundancy of $N(R-2)/R$ samples.  This alone suggests that the number of
rows should be maximized for best sensitivity.  However, when earth-rotation synthesis
is considered, the row length $N/R$ also becomes important.  By empirically comparing
redundancy metrics for \texttt{ln4x9}, \texttt{ln5x7}, and \texttt{ln6x6}, and
by including other comparisons for larger numbers of antennas,
we have determined that the
highest-redundancy configurations are generated by nearly square arrangements,
with $R\approx\sqrt{N}$.

In tuning the spacing between rows, there are several competing factors that need to be
considered.
The first is the increasing brightness of
galactic synchrotron emission at small $|\vec u|$
following a $C_\ell\propto\ell^{-3.7}$ scaling law \citep{chen2004}.
Secondly, the ability to control instrumental systematics decreases with low fringe-rates.  
The third is that
sensitivity reduces with increasing spacing owing to baselines moving more quickly through the $uv$-plane
as the earth rotates.  The fourth, which will be discussed in greater detail
in a forthcoming publication, is the fact that the frequency-dependence of $uv$-sampling
becomes increasingly problematic for foreground removal at longer baseline lengths.  Finally, the increasing
dominance of point-source foregrounds at higher $|\vec u|$ imply that there will be diminishing returns for
reducing foregrounds by increasing baseline length.  Taken together, these 
factors imply that the spacing between rows should target the shortest spacing at
which galactic synchrotron emission and/or instrumental systematics do not pose a problem.
To standardize the clustered antenna configurations we examine, 
we choose a 20-meter fiducial spacing between clusters.

Optimizing families of maximum-redundancy configuration styles is straight-forward,
algorithmically.  Automating a broader exploration of configuration space for maximally-redundant
configurations is much more difficult, owing to the extremely low entropy of these states in
configuration space.  Our experience has been that random processes are highly unlikely to 
encounter these configurations, even when a strong selective potential is applied.
In order to gain confidence that the manually-generated maximum-redundancy
configurations we explore are at least nearly optimal, it is
useful to compare them to the total redundancy of compact antenna
configurations (see \texttt{hex37} in Figures \ref{fig:config_study_configs}
and \ref{fig:rprofile}).  Figure \ref{fig:rprofile} illustrates that in
the $10<|\vec u|<20$ region for which it was optimized, the \texttt{ln5x7} 
configuration achieves approximately 50\% of the peak redundancy of \texttt{hex37}.
Hence, we may have confidence that while other configurations
might outperform \texttt{ln5x7}, they will not do so by more than a factor of two.  


Finally, it is worth pointing out that far from being a calibration liability, 
maximum-redundancy arrays may actually be more conducive to calibration 
than their minimum-redundancy counterparts.  
Redundant samplings of the $uv$-plane with many antenna pairings produce
independent measurements of the same quantity, facilitating the calibration
of per-antenna gain and phase parameters \citep{liu_et_al2010}.  Especially for
configurations involving shorter baselines sensitive to large-scale structures
on the sky,
the fact that many baselines fundamentally measure the same quantity can improve
calibration by easing the need for
an accurate sky model on which to base a self-calibration loop.

\subsection{Sensitivity Performance}
\label{sec:sensitivity_performance}

\begin{figure*}\centering
    \includegraphics[width=0.8\columnwidth]{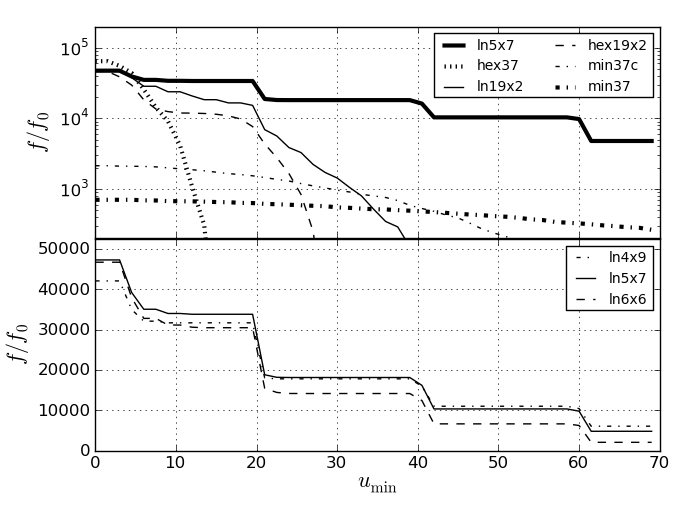}\caption{
Plotted above is the redundancy metric $f/f_0$ (relating to sensitivity, see
Eq. \ref{eq:redundancy_metric}) for several of the antenna configurations in Figure
\ref{fig:config_study_configs} as a function of the minimum distance $u_{\rm min}$ that may
be used for power-spectrum analysis, owing to
bright foreground emission and instrumental systematics associated with low fringe rates
(see \S\ref{sec:sensitivity_performance}).  The
most centrally condensed configuration (\texttt{hex37}) maximizes $f$ if regions within
$u_{\rm min}$
are not omitted.  If a region with radius $u_{\rm min}>5$ wavelengths is omitted,
configurations with larger separations are preferable, most notably
\texttt{ln5x7},
which dominates all other configurations out to twice the separation between antenna rows.
For comparison, two minimum-redundancy configurations 
(\texttt{min37} and \texttt{min37c}, see Figure \ref{fig:costas_minred}) are also plotted.
The lower plot highlights how
the number of rows in a configuration affects redundancy.
}\label{fig:rprofile}
\end{figure*}

Of the different configurations considered for maximum-redundancy arrays (see
Figure \ref{fig:config_study_configs}), the optimal for sensitivity choice
depends on the degree to which the shortest baselines are subject to
instrumental and celestial interference.  To parametrize our ignorance of what
are the shortest baselines that may be effectively used, we introduce a
parameter $u_{\rm min}$ to describe a minimum cutoff for baselines
contributing to power-spectrum sensitivity.  For the antenna configurations
shown in Figure \ref{fig:config_study_configs}, with corresponding
$uv$-coverage in Figure \ref{fig:config_study_uv}, we compute the redundancy
metric as a function of $u_{\rm min}$, omitting regions inside of $|\vec
u|<u_{\rm min}$ from the numerator in equation
\ref{eq:redundancy_metric}. The results are shown in Figure \ref{fig:rprofile}.

\begin{figure*}\centering
    \includegraphics[width=0.8\columnwidth]{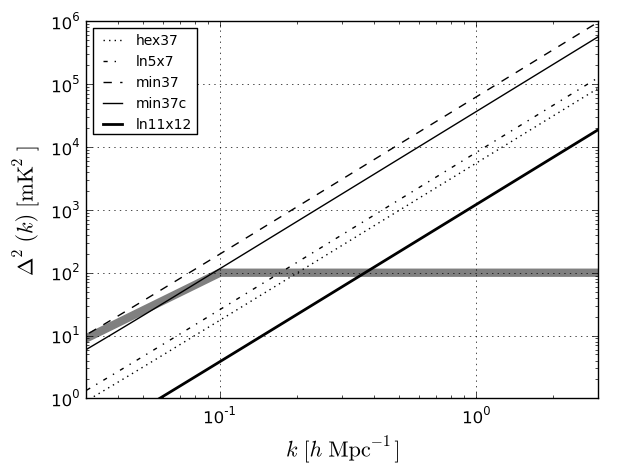}\caption{
This plot shows ideal, $1\sigma$ noise-sensitivity levels to the 21cm EoR
power spectrum using various array configurations at 150 MHz, assuming PAPER observing parameters
for 120-day
drift-scan observations over six sidereal hours of sky with a 6-MHz observing bandwidth (see
equation \ref{eq:multimode_redundant_sensitivity_fiducial}).
Except for \texttt{hex37} (where $u_{\rm min}=0$ is used), the sensitivities shown assume
$u_{\rm min}=10$ from Figure \ref{fig:rprofile}.  Minimum-redundancy configurations
(\texttt{min37} and \texttt{min37c}, above) show significantly reduced sensitivity relative to the best-performing
\texttt{ln5x7} maximum-redundancy configuration.  The \texttt{ln11x12} configuration is an extension
of the \texttt{ln5x7} design to 132 antennas, with a 3.75-meter pitch between antennas within
a row.  The thick black line denotes an optimistic toy model for peak 21cm EoR fluctuations (see
equation \ref{eq:model_eor_signal}).  
Predictions for the signal range between this curve and a factor of ten smaller.
In this plot, $k$ is dominated by
the line-of-sight component for the compact configurations we consider; at smaller $k$, sensitivity departs from a
power-law as the contribution of a baseline's length to $k$ becomes important.
    }\label{fig:config_study_sense}
\end{figure*}

If all baselines may be used effectively to measure $\Delta_{21}^2(k)$,
the most compact configuration (\texttt{hex32} in Figure \ref{fig:config_study_configs}) 
maximizes the redundancy metric.  If baselines shorter than five wavelengths are unusable,
however, the array configurations that are most effective
employ rows of antennas that have high instantaneous sampling redundancy and also generate substantial
redundancy through earth-rotation synthesis.  In particular, we show in the bottom panel
of Figure \ref{fig:rprofile} that the \texttt{ln5x7} configuration dominates all other configurations
within the $5<u_{\rm min}<20$ region for which its row spacing was tuned.  Based on the success
of this design, we extrapolate to a 132-antenna design consisting of 11 rows of 12 elements each, with
a 3.75-meter spacing between antennas within a row. 
This configuration (labeled \texttt{ln11x12} in Figure \ref{fig:config_study_sense})
was shown to have a higher redundancy metric than $8\times16$ and $16\times8$ designs.

Using the redundancy metric read from Figure \ref{fig:rprofile} at a chosen $u_{\rm min}$, we can
calculate a sensitivity as a function of $k$ by applying equation
\ref{eq:multimode_redundant_sensitivity_fiducial}.  In Figure \ref{fig:config_study_sense}, we plot the
sensitivities for selected configurations using observation parameters matched to the PAPER
experiment operating at 150 MHz with a 6-MHz observing bandwidth, assuming 120-day
drift-scan observations over six sidereal hours of sky using phase centers spaced two hours apart.
We compare these sensitivities with a toy reionization model derived from equation \ref{eq:model_eor_signal} 
under optimistic assumptions that produce $100~{\rm mK}^{2}$ peak fluctuations in the range 
$0.1<k<10h\ {\rm Mpc}^{-1}$.  As shown in Figure \ref{fig:config_study_sense}, we find that
a 132-antenna deployment of PAPER antennas will, under ideal conditions, have the requisite sensitivity to detect peak
21cm EoR fluctuations at a $3\sigma$ level at $k<0.25h\ {\rm Mpc}^{-1}$ with approximately four months of
observation.  This result is contingent upon antennas being deployed in a maximum-redundancy configurations,
and does not include the potential effects of foreground removal on sensitivity, which will be discussed in a future paper.
As shown by the sensitivity curves for minimum-redundancy configurations (\texttt{min37} and \texttt{min37c}
in Figure \ref{fig:config_study_sense}), maximum-redundancy configurations yield nearly an order-of-magnitude
improvement in sensitivity for 37-antenna arrays.  This improvement is larger for bigger arrays, owing
to the fact that $f/f_0\propto N^2$ for maximum-redundancy arrays, whereas
$f/f_0=1$ for minimum-redundancy arrays (except when scaling 
configurations below the critical size where adjacent samples of the $uv$-plane
are no longer independent; in this case, redundancy scales with maximum baseline length as
$f/f_0\propto u_{\rm max}^{-4}$).

Although these sensitivity figures were computed using observing parameters for the PAPER experiment, it
is straightforward to extrapolate them to other experiments.  In particular, equation 
\ref{eq:multimode_redundant_sensitivity_fiducial} makes very few assumptions about observing strategies
that may differ between experiments --- all such differences are grouped into the numerically computed
redundancy factor.  PAPER employs drift-scan observations that limit observing time toward
one phase center to approximately two hours.  For experiments that track the sky with dishes or
with station beam-forming, the amount of time spent observing toward the same phase center may be considerably
longer.  As a result, there may be additional redundancy generated by using earth-rotation synthesis 
over longer periods.  In all of these cases, computing the sampling generated for a single
phase center yields the correct redundancy value for use in equation 
\ref{eq:multimode_redundant_sensitivity_fiducial}.


A natural question that arises from the efficacy of antenna clustering for improving
sensitivity is whether, given the density of sampling within a row, it might be
desirable to employ larger antenna elements (perhaps parabolic cylinders) in lieu
of numerous smaller antenna elements.  Phrased differently, is the $\mathcal{O}(N)$ improvement to SNR
that results from implementing a fixed collecting area with $N$ antenna elements worth the $\mathcal{O}(N^2)$ cost
of correlating them?  In the limit that the correlator is a dominant cost in the construction
of an array, using large dishes, beam-forming antennas prior to correlation, or even
operating separate sub-arrays may all represent attractive options for more cheaply improving
sensitivity.
On the other hand, extrapolating from currently deployed systems using Moore's Law applied to 
computational density suggests correlators might
not be the dominant cost for forthcoming arrays (Jason Manley, personal
communication), in which case 
smaller elements yield the best sensitivity for a fixed collecting area.  This case may be even stronger,
noting that array configurations consisting of parallel rows are particularly
conducive to correlation via electric-field gridding techniques \citep{tegmark_zaldarriaga2009}
that replace the $\mathcal{O}(N^2)$ scaling of the computational cost of correlators with an $\mathcal{O}(N\ {\rm log}N)$
scaling based on the fast Fourier transform algorithm.

\section{Conclusion}
\label{sec:conclusion}

Reionization experiments aiming to detect the power spectrum of 
21cm EoR fluctuations will need to achieve a tremendous level of
foreground removal.  For characterizing these foregrounds,
minimum-redundancy array configurations are most useful.  However,
as efforts turn to constraining the three-dimensional
power spectrum of 21cm EoR fluctuations, the maximum-redundancy configurations
we have presented provide a substantial increase in sensitivity over
their minimum-redundancy counterparts in the regime where sensitivity is not limited
by sample variance in the cosmological signal --- a regime that 21cm
reionization arrays will find themselves in for the foreseeable future.
While the most compact antenna arrangements
yield the highest-redundancy sampling of the $uv$-plane, the performance of these arrangements
may be compromised by increased foreground brightness at low $|\vec u|$ and by
instrumental systematics associated with low fringe-rates.  

In order to avoid the difficulties of working near the origin of the $uv$-plane, 
we show how maximum-redundancy array configurations can be tuned to
regions of the $uv$-plane where foregrounds and systematics are likely to be smaller.  Of particular interest
are a class of configurations using parallel rows of antennas that generate substantial
instantaneous sampling redundancy, but are also aligned to enhance redundancy through
earth-rotation synthesis.  Using such a configuration, we demonstrate
that under ideal conditions, a 132-antenna deployment of PAPER observing for 120 days will have the requisite
sensitivity to detect the power spectrum of the brightest expected reionization fluctuations 
at a $3\sigma$ level at $k<0.25h\ {\rm Mpc}^{-1}$, using bins of $\Delta{\rm ln}~k=1$.  
The real-world sensitivity of such an array will be affected by foreground removal requirements and actual
performance, and could be worse than this.

Next-generation 21cm reionization arrays such as phase-II of HERA
will have improved sensitivity.
Even so, HERA's choice of array configuration must balance the competing
goals of imaging bright EoR structures and characterizing the power-spectrum of EoR
fluctuations.  For current and future arrays, this choice must be informed by parameters that are currently
poorly constrained: the degree to which foregrounds can be modeled, removed, or otherwise differentiated
from the 21cm EoR signal; the angular power spectra of the dominant foregrounds; 
the nature of instrumental systematics that arise; and the geometry and collecting area of the most
effective antenna elements.  Exploration of these design parameters is underway with phase-I HERA 
efforts such as PAPER and the MWA.
PAPER is in a unique
position to use the mobility of its antennas to explore different
configurations for 21cm reionization work.
Maximum-redundancy arrays can be used to push sensitivity limits for
power-spectrum measurements while minimum-redundancy configurations
will help glean more information about foreground properties.  
Near-term activities can explore the results of tuning
array sensitivity relative to foreground brightness and examine the influence
of cross-coupling and crosstalk on power-spectrum measurements.  Continued work 
in this area will aim to establish an optimal array design for
next-generation 21cm EoR work.

\section*{Acknowledgement}

We dedicate this paper to the memory of Don Backer, who pioneered PAPER
and laid the foundations for HERA.
AP thanks Philip Matchett Wood for helpful discussions of Golomb rulers and Costas arrays
and acknowledges support from the NSF AAPF (\#0901961) and
from the Charles H. Townes Postdoctoral Fellowship.  MM acknowledges
support form the Einstein Postdoctoral Fellowship.  PAPER is supported
through the NSF-AST program (\#0804508). 

\appendix
\section{Sensitivity from Combining Independent $\vec k$-modes}
\label{app:multimode_sensitivity}

This appendix derives equation \ref{eq:multimode_sensitivity_fiducial} 
in full.
We treat each of the three sources of independent samples mentioned 
in \S\ref{sec:sense_independent_modes} separately.
We begin with equation \ref{eq:sensitivity_fiducial}, reproduced here:
\begin{align}
    \Delta_{\rm N}^2(k)\approx &~2.8\times10^4
    \left[\frac{k}{0.1h\ {\rm Mpc}^{-1}}\right]^3
    \left[\frac{\Omega}{0.76\ {\rm sr}}\right]^\frac32
    \nonumber\\&\times
    \left[\frac{T_{\rm sys}}{500\ {\rm K}}\right]^2
    \left[\frac{120\ {\rm days}}{t_{\rm days}}\right] 
    \left[\frac{|\vec u|}{20}\right]
    \ {\rm mK}^2.
    \label{app_eq:sensitivity_fiducial}
\end{align}

\subsection{Combining Modes Along the Line-of-Sight}
\label{app:combining_los_modes}
Let us bin the line-of-sight modes in
logarithmic increments.  For a fixed logarithmic bin size of $\Delta\ln k$, the number of modes in 
each bin grows linearly with $k$ and linearly with the chosen observing bandwidth $B$.
The bandwidth term
enters because it sets the total number of $k$-modes measured for a constant frequency
channel resolution.  We incorporate these scalings into equation 
\ref{app_eq:sensitivity_fiducial}, using that sensitivity will scale as the 
square-root of the number of independent $\vec k$-modes binned, and counting the modes in a 
fiducial bin to set the prefactor.
For a bin centered around $k = 0.1h\ {\rm Mpc^{-1}}$, a bin of width $\Delta \ln k$ has bin
edges at $0.06h$ and $0.165h\ {\rm Mpc^{-1}}$.  A bandwidth of 6MHz produces a $\vec k$-space resolution
of $2\pi/YB \approx 0.083h\ {\rm Mpc^{-1}}$, where $Y$ is given by equation \ref{eq:rcom_los}.
Therefore, we count approximately $1.27$ independent $\vec k$-modes, resulting in a 
$\sqrt{1.27} \approx 1.13$-fold increase in sensitivity, or
\begin{align}
    \Delta_{\rm N}^2(k)\approx &~2.48\times10^4
    \left[\frac{k}{0.1h\ {\rm Mpc}^{-1}}\right]^\frac52
    \left[\frac{6\ {\rm MHz}}{B}\right]^\frac12
    \left[\frac1{\Delta{\rm ln}~k}\right]^\frac12
    \nonumber\\&\times
    \left[\frac{\Omega}{0.76\ {\rm sr}}\right]^\frac32
    \left[\frac{T_{\rm sys}}{500\ {\rm K}}\right]^2
    \left[\frac{120\ {\rm days}}{t_{\rm days}}\right] 
    \left[\frac{|\vec u|}{20}\right]
    \ {\rm mK}^2.
    \label{app_eq:sensitivity_los_bins}
\end{align}

\subsection{Combining Time Samples and Modes Across the $uv$-plane}
To combine modes measured by different baselines throughout the $uv$-plane
and to calculate the amount of time a baseline samples a single $(u,v,\eta)$ mode,
we must assume an array configuration.  
As stated in \S\ref{sec:sense_independent_modes}, we assume an array configuration
that generates uniform, non-overlapping coverage in the $uv$-plane out to a radius $u_{\rm max}$.  
This assumption makes the problem
algebraically tractable, and is similar to
the minimum-redundancy arrays discussed in \S\ref{sec:minimum_redundancy}.  We explicitly
sum measurements from $(u,v,\eta)$-mode bins, or ``$uv$-pixels'', over the $uv$-plane.
We use the calculated the noise in any 
$uv$-pixel from equation \ref{app_eq:sensitivity_los_bins}, and add up all the samples within
each ring of constant $uv$-distance $|\vec u|$.  Finally, we sum over all the rings out to $u_{\rm max}$.

Let us define two additional terms.  Let the sampling density of the $uv$-plane,
$\rho$, be given by
\begin{equation}
\label{app_eq:def_rho}
\rho\equiv \frac{N_{\rm bl}}{\pi u_{\rm max}^2} \approx \frac{N^2}{2\pi u_{\rm max}^2},
\end{equation}
where $N_{\rm bl}$ is the number of baselines and $N$ is the number of antenna elements.  This
equation restates our assumption that the $uv$-plane is uniformly sampled out to a 
radius $u_{\rm max}$.  We also define $t_{\rm per\_mode}$, the time a baseline samples a single $uv$-pixel
relative to a chosen phase center before earth rotation moves it into another pixel:
\begin{equation}
t_{\rm per\_mode}\equiv t_{20} \left[\frac{\Omega_0}{\Omega}\right]^\frac12 \left[\frac{20}{|\vec u|} \right],
\end{equation}
where $t_{20}$ is the amount of time a 20-wavelength baseline spends in one pixel (used a fiducial 
scale), $\Omega_0$ is a fiducial primary beam size (0.76 sr for PAPER),
 and $|\vec u|$ is the baseline length.  Note that, per the assumptions of the derivation
in \S\ref{sec:sense_independent_modes}, we neglect here the possibility that multiple baselines may sample the
same $uv$-pixel under earth-rotation synthesis.  Note also
that $t_{20}$ depends on the array latitude; for PAPER it is approximately 13 minutes; this factor 
has already been absorbed into the prefactor in equation \ref{eq:sensitivity_fiducial}.  

With these terms defined, we first calculate the number of $uv$-pixels within a ring of radius $u$.
We choose an arbitrary ring width $w$, which will drop out of the derivation later.  The number of baselines
that sample within this ring is then:
\begin{equation}
N_{\rm bl} \approx 2\pi |\vec u|w\rho.
\end{equation}
The number of pixels sampled depends on the observing time, $t_{\rm per\_day}$ and the time spent in
each pixel, $t_{\rm per\_mode}$:
\begin{equation}
N_{\rm px,ring} = N_{\rm bl} \left[\frac{t_{\rm per\_day}}{t_{\rm per\_mode}} \right].
\end{equation}
To calculate the sensitivity of one ring in the $uv$-plane, we average over the sensitivity of
each pixel within the ring.  Each pixel within the ring has equal sensitivity
(e.g. equation \ref{app_eq:sensitivity_los_bins}),
so this is a simple, unweighted average:
\begin{equation}
\Delta^2_{\rm N,ring} = \frac{\sum\limits_{\rm px}\Delta^2_{\rm N,px}}{N_{\rm px}} = \frac{\Delta^2_{\rm N,px}}{\sqrt{N_{\rm px}}},
\end{equation}
where $\Delta^2_{\rm N,px}$ is given by equation \ref{app_eq:sensitivity_los_bins}.
Plugging in values calculated above gives:
\begin{equation}
\label{app_eq:delta2_in_ring}
\Delta^2_{\rm N,ring} = \frac{\Delta^2_{\rm N,px}}{\sqrt{2\pi |\vec u|w\rho\frac{t_{\rm per\_day}}{t_{\rm per\_mode}}}} = \frac{\Delta^2_{\rm N,px}}{\sqrt{2\pi |\vec u|w\rho\frac{t_{\rm per\_day}}{t_{20}}\left[\frac{\Omega}{\Omega_0}\right]^\frac12\frac{|\vec u|}{20}}}.
\end{equation}

The next step is to combine all the measurements from different rings.  In the case presented here,
the noise power in each ring is equal, as the $\sqrt{|\vec u|^2}$ term in the denominator
of equation \ref{app_eq:delta2_in_ring} cancels the $|\vec u|$ term in equation 
\ref{app_eq:sensitivity_los_bins}.  We can therefore do another unweighted average to combine 
the rings:
\begin{equation}
\Delta^2_{\rm N} = \frac{\sum\limits_{\rm rings}\Delta^2_{\rm N,ring}}{N_{\rm rings}} = \frac{\Delta^2_{\rm N,ring}}{\sqrt{N_{\rm rings}}} = \frac{\Delta^2_{\rm N,ring}}{\sqrt{u_{\rm max}/w}},
\end{equation}
where the last step uses the fact that the number of rings is the radius of the circle over the width
of a ring, $u_{\rm max}/w$.  Plugging equations \ref{app_eq:sensitivity_los_bins}
and \ref{app_eq:def_rho}
into this equation yields:
\begin{equation}
\Delta^2_{\rm N} = \Delta^2_{\rm N,px} \left[\frac{20}{|\vec u|}\right] \left[\frac{\Omega_0}{\Omega}\right]^\frac14 \left[\frac{t_{20}}{t_{\rm per\_day}}\right]^\frac12 \left[\frac{u_{\rm max}}{20}\right]^\frac12 \left[\frac{1}{N}\right].  
\end{equation}
We then reach our final result by substituting in equation \ref{app_eq:delta2_in_ring},
choosing a fiducial observation time of $t_{\rm per\_day}=6$ hours (recall that $t_{20}\approx 13$min), and
choosing an array size of $N=32$ elements:
\begin{align}
    \Delta_{\rm N}^2(k)\approx &~150   
    \left[\frac{k}{0.1h\ {\rm Mpc}^{-1}}\right]^\frac52
    \left[\frac{6\ {\rm MHz}}{B}\right]^\frac12
    \left[\frac1{\Delta{\rm ln}~k}\right]^\frac12
    \nonumber\\&\times
    \left[\frac{\Omega}{0.76\ {\rm sr}}\right]^\frac54
    \left[\frac{T_{\rm sys}}{500\ {\rm K}}\right]^2
    \left[\frac{6\ {\rm hrs}}{t_{\rm per\_day}}\right]^\frac12
    \nonumber\\&\times
    \left[\frac{120\ {\rm days}}{t_{\rm days}}\right]
    \left[\frac{u_{\rm max}}{20}\right]^{\frac12}
    \left[\frac{32}{N}\right]
    \ {\rm mK}^2.
    \label{app_eq:multimode_sensitivity_fiducial}
\end{align}

\section{Sensitivity from Combining Redundant Samples of $\vec k$-modes}
\label{app:multimode_redundant_sensitivity}

In this appendix, we first motivate our definition of the redundancy factor presented 
in equation \ref{eq:redundancy_metric}
in \ref{app:redund_metric}.  Next, in \ref{app:maxred_sense_deriv} we derive equation \ref{eq:multimode_redundant_sensitivity_fiducial}
in full.  Finally, we
present an analytic calculation of the redundancy metric for a filled circular aperture in \ref{app:circ_ap_deriv}.

Several times below we will refer to sensitivity 
relative to a fiducial measurement.  
For this benchmark, we choose a single baseline with a one-second integration, which we call
$\Delta_{\rm {N,1sec}}^2$.
The choice of fiducial integration time is arbitrary, but affects the scaling constants in
the equations that follow.  For baselines with length $|\vec u|<10^4$ 
(essentially all baselines useful for EoR measurements), earth-rotation is unimportant
on one-second timescales.  Therefore, our fiducial measurement is equal to equation 
\ref{eq:sensitivity},
\begin{equation}
\Delta_{\rm N,1sec}^2(k)\equiv X^2Y\frac{k^3}{2\pi^2}\frac{\Omega}{2t}T_{\rm sys}^2,
\label{eq:delta_1sec}
\end{equation}
with $t=1$ second. 

\subsection{Motivation for the Redundancy Metric}
\label{app:redund_metric}
In this section, we outline a formalism for computing the sensitivity of a generic 
antenna array, expressed in terms of a redundancy metric.
The sensitivity calculation in \S\ref{sec:sense_independent_modes} and Appendix \ref{app:multimode_sensitivity}
assumed a $uv$-coverage that produces equal sensitivity in
any ring of the $uv$-plane.  We were then able to perform an unweighted average to get the final
sensitivity by summing over rings.  More generally, the final sensitivity will be a weighted
average of the sensitivity of all the $uv$-pixels:
\begin{equation}
\label{eq:sense_weighted_average}
\Delta_{\rm N}^2 = \frac{\sum\limits_{i}w_{i}\Delta_{\rm{N},i}^2}{\sum\limits_{i}w_{i}},
\end{equation}
where $i$ is an index labeling an individual $uv$-pixel and $\Delta_{\rm{N},i}^2$ is the noise
variance of a mode.
The optimal weights for any pixel are proportional to the inverse variance of noise in that pixel. 
Since
repeated measurements of a $uv$-pixel add coherently in temperature, 
redundant measurements beat down the noise in temperature-squared \emph{linearly} 
with the number of samples.  Therefore, we choose optimal inverse-variance
weights $w_i=n_i^2$,
where $n_i$ is the number of fiducial samples in
that pixel, and equation \ref{eq:sense_weighted_average} then becomes:
\begin{equation}
\Delta_{\rm N}^2 = \frac{\sum\limits_{i}n^2_i\frac{\Delta_{\rm N,1sec}^2}{n_i}}{\sum\limits_{i}n^2_i} = \frac{\sum\limits_{i}n_i\Delta_{\rm N,1sec}^2}{\sum\limits_{i}n^2_i}
\end{equation}
We further simplify this equation by noting that the coefficients of a weighted sum 
of random numbers will add in quadrature:
\begin{equation}
\label{eq:sum_over_red_px}
\Delta_{\rm N}^2 = \frac{\sqrt{\sum\limits_{i}n_i^2\Delta_{\rm N,1sec}^4}}{\sum\limits_{i}n^2_i} = \frac{\Delta_{\rm N,1sec}^2}{\sqrt{\sum\limits_{i}n_i^2}},
\end{equation}

We compare this result to the reduction in noise that would occur without any two samples being
of the same mode (i.e. $n_i$ being uniformly unity):
\begin{equation}
\Delta_{\rm N,no-red}^2 = \frac{\sqrt{\sum\limits_{j}\Delta_{\rm N,1sec}^4}}{\sum\limits_{j}1} = \frac{\Delta_{\rm N,1sec}^2}{{\sqrt{\sum\limits_{j}1}}}
\end{equation}
where the index $j$ labels pixels.  (We use a different letter here to differentiate these
modes from those used in equation \ref{eq:sum_over_red_px}).  The number of samples has remained constant between
this case and the one calculated above, $\sum\limits_{j}1 = {\sum\limits_{i}n_i}$, giving us:
\begin{equation}
\Delta_{\rm N,no-red}^2 = \frac{\Delta_{\rm N,1sec}^2}{\sqrt{\sum\limits_{i}n_i}}
\end{equation}

The relative improvement of redundant sampling over the completely non-redundant case
is the ratio between these two terms:
\begin{equation}
\frac{\Delta_{\rm N}^2}{\Delta_{\rm N,no-red}^2} = \frac{\Delta_{\rm N,1sec}^2}{\sqrt{\sum\limits_{i}n_i^2}}\frac{{\sqrt{\sum\limits_{i}n_i}}}{\Delta_{\rm N,1sec}^2} = \sqrt{\frac{\sum\limits_{i}n_i}{\sum\limits_{i}n_i^2}}
\end{equation}
We use this result to motivate the definition of our metric for the sampling redundancy
given in equation \ref{eq:redundancy_metric}:
\begin{equation}
    \frac{f}{f_0}\equiv\frac{\sum_i{n_i^2}}{\sum_i{n_i}}.
\label{app_eq:redundancy_metric}
\end{equation}.

\subsection{Derivation of Maximum-Redundancy Sensitivity}
\label{app:maxred_sense_deriv}

To derive array sensitivity including the effects of sampling redundancy, we
begin by evaluating equation \ref{eq:delta_1sec} at $z=9$, substituting in equation
\ref{eq:x2y_vs_z}:
\begin{equation}
\Delta_{\rm N,1sec}^2(k) \approx \frac{540.}{4\pi^2}k^3\Omega T_{\rm sys}^2 \frac{h^{-3}\rm Mpc^3}{\rm sr~Hz~s}.
\end{equation}
Expressing the
result in terms of our fiducial observing and telescope parameters gives:
\begin{align}
    \Delta_{\rm N,1sec}^2(k)\approx &~2.6\times10^9
    \left[\frac{k}{0.1h\ {\rm Mpc}^{-1}}\right]^3
    \left[\frac{\Omega}{0.76\ {\rm sr}}\right]^\frac32
    \left[\frac{T_{\rm sys}}{500\ {\rm K}}\right]^2
    {\rm mK}^2.
\end{align}
Next, we incorporate several sensitivity contributions calculated previously: a factor
of 1.13 for logarithmic binning of 
line-of-sight modes (derived in Appendix \ref{app:combining_los_modes}), a factor of $(2.16 \times
10^4)^\frac12$ for the number of independent 1-second observations
in a 6-hour observing window, and a factor of 120 for
the number of days observed.
There is also a factor of $(N_{\rm baselines})^\frac12$ sensitivity increase,
since each baseline provides an independent sample at every integration.  For
our fiducial array of 32 antennas, this term is $\approx\sqrt{512}$.  The result is an
expression for the sensitivity of an array, assuming that every integration of every baseline is
treated as a sample of an independent $\vec k$-mode:
\begin{align}
    \Delta_{\rm N}^2(k)\approx &~5.8\times10^3
    \left[\frac{k}{0.1h\ {\rm Mpc}^{-1}}\right]^\frac52
    \left[\frac{6\ {\rm MHz}}{B}\right]^\frac12
    \left[\frac1{\Delta{\rm ln}~k}\right]^\frac12
    \nonumber\\&\times
    \left[\frac{\Omega}{0.76\ {\rm sr}}\right]^\frac32
    \left[\frac{T_{\rm sys}}{500\ {\rm K}}\right]^2
    \left[\frac{6\ {\rm hrs}}{t_{\rm per\_day}}\right]^\frac12
    \left[\frac{120\ {\rm days}}{t_{\rm days}}\right] 
    \ {\rm mK}^2.
\end{align}
Finally, as derived in Appendix \ref{app:redund_metric}, the effects of sampling
redundancy may be included by introducing a factor of $[f/f_0]^\frac12$.
Using a fiducial value of $f/f_0 = 10^4$, we have our final result:
\begin{align}
    \Delta_{\rm N}^2(k)\approx &~58
    \left[\frac{k}{0.1h\ {\rm Mpc}^{-1}}\right]^\frac52
    \left[\frac{6\ {\rm MHz}}{B}\right]^\frac12
    \left[\frac1{\Delta{\rm ln}~k}\right]^\frac12
    \nonumber\\&\times
    \left[\frac{\Omega}{0.76\ {\rm str}}\right]
    \left[\frac{T_{\rm sys}}{500\ {\rm K}}\right]^2
    \left[\frac{6\ {\rm hrs}}{t_{\rm per\_day}}\right]^\frac12
    \nonumber\\&\times
    \left[\frac{120\ {\rm days}}{t}\right]
    \left[\frac{32}{N}\right]
    \left[\frac{10^4 f_0}{f}\right]^\frac12
    \ {\rm mK}^2.
\end{align}

\subsection{Calculating $f/f_0$ for a Filled Circular Aperture}
\label{app:circ_ap_deriv}
Here we explicitly calculate the instantaneous redundancy of an array where
antennas are arranged to uniformly sample an aperture within a region defined by 
$|\vec r|<R$, with zero sampling elsewhere.  Using that the convolution of two disks of area
$\pi R^2$ is a cone of height $h=\pi R^2$ and base radius $2R$, we can compute:
\begin{align}
    f&=\frac{\int_0^{2R}{2\pi r\ dr\ h^2\left(1-\frac{r}{2R}\right)^2}}
            {\int_0^{2R}{2\pi r\ dr\ h\left(1-\frac{r}{2R}\right)}}\nonumber\\
     &=\frac{\phantom{^\Big[}2\pi h^2\left(\frac{r^2}{2} - 2\frac{r^3}{6R} + \frac{r^4}{16R^2}\right)\big|_0^{2R}}
            {2\pi h\left(\frac{r^2}{2} - \frac{r^3}{6R}\right)\big|_0^{2R}}
    =\frac{h}{2}
    \label{eq:f_disk}
\end{align}
Using that $h=\pi R^2=NA_e$, where $N$ is the number of antennas and $A_e$ is the effective area of a
single antenna, we have $f=NA_e/2=Nf_0$.  
In general, the
redundancy metric must be calculated numerically to account for more complicated
array configurations and the effects of earth-rotation synthesis.
%
\bibliographystyle{apj}
\bibliography{biblio}

\begin{thebibliography}{39}
\expandafter\ifx\csname natexlab\endcsname\relax\def\natexlab#1{#1}\fi

\bibitem[{Babcock(1953)}]{babcock1953}
Babcock, W. 1953, Bell System Technical Journal, 31, 63

\bibitem[{{Bernardi} {et~al.}(2010){Bernardi}, {de Bruyn}, {Harker},
  {Brentjens}, {Ciardi}, {Jeli{\'c}}, {Koopmans}, {Labropoulos}, {Offringa},
  {Pandey}, {Schaye}, {Thomas}, {Yatawatta}, \& {Zaroubi}}]{bernardi_et_al2010}
{Bernardi}, G., et~al. 2010, \aap, 522, A67+

\bibitem[{{Bowman} {et~al.}(2006){Bowman}, {Morales}, \&
  {Hewitt}}]{bowman_et_al2006}
{Bowman}, J., et al. 2006, \apj, 638, 20

\bibitem[{{Bowman} {et~al.}(2009){Bowman}, {Morales}, \&
  {Hewitt}}]{bowman_et_al2009}
---. 2009, \apj, 695, 183

\bibitem[{{Bowman} \& {Rogers}(2010)}]{bowman_rogers2010}
{Bowman}, J. \& {Rogers}, A. 2010, \nat, 468, 796

\bibitem[{{Chen}(2004)}]{chen2004}
{Chen}, X. 2004, arXiv:astro-ph/0409733

\bibitem[{{Clark}(1999)}]{clark1999}
{Clark}, B. 1999, in ASP Conf. Series,
  Vol. 180, Synth. Imaging in Radio Ast. II, ed. {Taylor,
  Carilli, \& Perley}, 1--+

\bibitem[{{{Comm. for a Decadal Survey of A\&A; NRC}}(2010)}]{astro2010}
{{Comm. for a Decadal Survey of A\&A; NRC}}. 2010, New Worlds, New Horizons in
  A\&A (Natl. Acad. Press)

\bibitem[{{Cornwell} {et~al.}(2003){Cornwell}, {Golap}, \&
  {Bhatnagar}}]{cornwell_et_al2003}
{Cornwell}, T., et al. 2003, {W-Projection: A New
  Algorithm for Non-Coplanar Baselines}, EVLA Memo~67

\bibitem[{Costas(1984)}]{costas1984}
Costas, J. 1984, in Proc. of the IEEE, Vol.~72, 996--1009

\bibitem[{{de Villiers}(2007)}]{devilliers2007}
{de Villiers}, M. 2007, \aap, 469, 793

\bibitem[{{Field}(1958)}]{field1958}
{Field}, G. 1958, Proc. IRE, 46, 240

\bibitem[{{Furlanetto} {et~al.}(2006){Furlanetto}, {Oh}, \&
  {Briggs}}]{furlanetto_et_al2006}
{Furlanetto}, S., et al. 2006, \physrep, 433, 181

\bibitem[{Golomb \& Taylor(1984)}]{golomb_taylor1984}
Golomb, S. \& Taylor, H. 1984, Proc. of the IEEE, 72, 1143

\bibitem[{Golomb \& Gong(2004)}]{golomb_gong2004}
Golomb, S. \& Gong, G. 2004, Signal Design for Good Correlation: For
  Wireless Comm., Crypt., and Radar (NY, NY: Cambridge
  Univ. Press)

\bibitem[{{Jeli{\'c}} {et~al.}(2008){Jeli{\'c}}, {Zaroubi}, {Labropoulos},
  {Thomas}, {Bernardi}, {Brentjens}, {de Bruyn}, {Ciardi}, {Harker},
  {Koopmans}, {Pandey}, {Schaye}, \& {Yatawatta}}]{jelic_et_al2008}
{Jeli{\'c}}, V., et~al.  2008, \mnras, 389, 1319

\bibitem[{{Keto}(1997)}]{keto1997}
{Keto}, E. 1997, \apj, 475, 843

\bibitem[{{Lidz} {et~al.}(2008){Lidz}, {Zahn}, {McQuinn}, {Zaldarriaga}, \&
  {Hernquist}}]{lidz_et_al2008}
{Lidz}, A., et~al.  2008, \apj, 680, 962

\bibitem[{{Liu} {et~al.}(2010){Liu}, {Tegmark}, {Morrison}, {Lutomirski}, \&
  {Zaldarriaga}}]{liu_et_al2010}
{Liu}, A., el~al. 2010, \mnras, 408, 1029

\bibitem[{{Lonsdale} {et~al.}(2009){Lonsdale}, {Cappallo}, {Morales}, {Briggs},
  {Benkevitch}, {Bowman}, {Bunton}, {Burns}, {Corey}, {Desouza}, {Doeleman},
  {Derome}, {Deshpande}, {Gopala}, {Greenhill}, {Herne}, {Hewitt}, {Kamini},
  {Kasper}, {Kincaid}, {Kocz}, {Kowald}, {Kratzenberg}, {Kumar}, {Lynch},
  {Madhavi}, {Matejek}, {Mitchell}, {Morgan}, {Oberoi}, {Ord},
  {Pathikulangara}, {Prabu}, {Rogers}, {Roshi}, {Salah}, {Sault}, {Shankar},
  {Srivani}, {Stevens}, {Tingay}, {Vaccarella}, {Waterson}, {Wayth}, {Webster},
  {Whitney}, {Williams}, \& {Williams}}]{lonsdale_et_al2009}
{Lonsdale}, C., et~al. 2009, Proc. of the IEEE, 97, 1497

\bibitem[{{Madau} {et~al.}(1997){Madau}, {Meiksin}, \&
  {Rees}}]{madau_et_al1997}
{Madau}, P., et al. 1997, \apj, 475, 429

\bibitem[{{McQuinn} {et~al.}(2007){McQuinn}, {Lidz}, {Zahn}, {Dutta},
  {Hernquist}, \& {Zaldarriaga}}]{mcquinn07}
{McQuinn}, M., et~al. 2007, \mnras, 377, 1043

\bibitem[{{McQuinn} {et~al.}(2006){McQuinn}, {Zahn}, {Zaldarriaga},
  {Hernquist}, \& {Furlanetto}}]{mcquinn_et_al2006}
{McQuinn}, M., et~al. 2006, \apj, 653, 815

\bibitem[{{Morales}(2005)}]{morales2005}
{Morales}, M. 2005, \apj, 619, 678

\bibitem[{{Morales} \& {Hewitt}(2004)}]{morales_hewitt2004}
{Morales}, M. \& {Hewitt}, J. 2004, \apj, 615, 7

\bibitem[{{Paciga} {et~al.}(2010){Paciga}, {Chang}, {Gupta}, {Nityanada},
  {Odegova}, {Pen}, {Peterson}, {Roy}, \& {Sigurdson}}]{paciga_et_al2010}
{Paciga}, G., et~al. 2010, ArXiv: 1006.1351

\bibitem[{{Park} {et~al.}(2003){Park}, {Ng}, {Park}, {Liu}, \&
  {Umetsu}}]{park_et_al2003}
{Park}, C., et~al. 2003, \apj, 589, 67

\bibitem[{{Parsons} {et~al.}(2010){Parsons}, {Backer}, {Foster}, {Wright},
  {Bradley}, {Gugliucci}, {Parashare}, {Benoit}, {Aguirre}, {Jacobs},
  {Carilli}, {Herne}, {Lynch}, {Manley}, \& {Werthimer}}]{parsons_et_al2010}
{Parsons}, A., et~al. 2010, \aj, 139, 1468

\bibitem[{{Pen} {et~al.}(2009){Pen}, {Chang}, {Hirata}, {Peterson}, {Roy},
  {Gupta}, {Odegova}, \& {Sigurdson}}]{pen_et_al2009}
{Pen}, U., et~al. 2009, \mnras, 399, 181

\bibitem[{{Petrovic} \& {Oh}(2010)}]{petrovic_oh2010}
{Petrovic}, N. \& {Oh}, S. 2010, ArXiv: 1010.4109

\bibitem[{Robinson(1985)}]{robinson1985}
Robinson, J. 1985, IEEE Trans. on Information Theory, 31, 781

\bibitem[{{Rottgering} {et~al.}(2006){Rottgering}, {Braun}, {Barthel}, {van
  Haarlem}, {Miley}, {Morganti}, {Snellen}, {Falcke}, {de Bruyn}, {Stappers},
  {Boland}, {Butcher}, {de Geus}, {Koopmans}, {Fender}, {Kuijpers},
  {Schilizzi}, {Vogt}, {Wijers}, {Wise}, {Brouw}, {Hamaker}, {Noordam},
  {Oosterloo}, {Bahren}, {Brentjens}, {Wijnholds}, {Bregman}, {van Cappellen},
  {Gunst}, {Kant}, {Reitsma}, {van der Schaaf}, \& {de
  Vos}}]{rottgering_et_al2006}
{Rottgering}, H., et~al. 2006, arXiv:astro-ph/0610596

\bibitem[{{Santos} {et~al.}(2005){Santos}, {Cooray}, \&
  {Knox}}]{santos_et_al2005}
{Santos}, M., et al. 2005, \apj, 625, 575

\bibitem[{Sidon(1932)}]{sidon1932}
Sidon, S. 1932, Mathematische Annalen, 106, 536

\bibitem[{{Tegmark} \& {Zaldarriaga}(2009)}]{tegmark_zaldarriaga2009}
{Tegmark}, M. \& {Zaldarriaga}, M. 2009, \prd, 79, 083530

\bibitem[{{Thompson}(1999)}]{thompson1999}
{Thompson}, A. 1999, in ASP Conf.
  Series, Vol. 180, Synth. Imaging in Radio Ast. II, ed. {Taylor,
  Carilli, \& Perley}, 11--+

\bibitem[{{Trac} \& {Cen}(2007)}]{trac07}
{Trac}, H. \& {Cen}, R. 2007, \apj, 671, 1

\bibitem[{{Wrobel} \& {Walker}(1999)}]{wrobel_walker1999}
{Wrobel}, J. \& {Walker}, R. 1999, in ASP
  Conf. Series, Vol. 180, Synth. Imaging in Radio Ast. II, ed.
  {Taylor}, {Carilli}, \& {Perley}, 171--+

\bibitem[{{Zahn} {et~al.}(2007){Zahn}, {Lidz}, {McQuinn}, {Dutta}, {Hernquist},
  {Zaldarriaga}, \& {Furlanetto}}]{zahn_et_al2007}
{Zahn}, O., et~al. 2007, \apj, 654, 12

\end{thebibliography}


\end{document}